\documentclass[aps,ams,amsmath,prx,longbibliography,twocolumn]{revtex4-1}
\usepackage{graphics}
\usepackage{graphicx}
\usepackage{epstopdf,epsfig}
\usepackage{amsmath}
\usepackage{amssymb}
\usepackage{amsfonts}
\usepackage{mathrsfs}
\usepackage{bm}
\usepackage{xcolor}
\usepackage{bbm}
\usepackage{hyperref}
\usepackage[normalem]{ulem}
\usepackage{comment}
\usepackage{soul}
\usepackage[varg]{txfonts}

\newcommand{\dd}{\mathrm{d}}
\newcommand{\im}{\mathrm{Im}}
\newcommand{\re}{\mathrm{Re}}

\begin{document}

\title{Hydrodynamic fluctuations in Luttinger liquids}

\author{Alex Levchenko}
\affiliation{Department of Physics, University of Wisconsin--Madison, Madison, Wisconsin 53706, USA}

\date{October 27, 2026}

\begin{abstract}
We develop the theory of hydrodynamic fluctuations for one-dimensional quantum liquids. At low temperatures the relaxation of interacting electrons in a single-channel wire separates into two classes: equilibration, which changes the numbers of right and left movers through backscattering and is exponentially slow, and thermalization, which redistributes energy and momentum at a fixed number of movers at a power-law rate. This separation of time scales gives the electron liquid a two-fluid description analogous to superfluid $^4$He, with the zero mode of the Luttinger liquid as the superfluid component and the gas of thermal excitations as the normal component. Using Onsager reciprocity and the fluctuation-dissipation theorem, we add Langevin fluxes to the two-fluid hydrodynamic equations, including a mutual-friction noise generated by rare backscattering events that has no counterpart in helium. We apply the theory to Coulomb-coupled quantum wires and calculate the dynamic structure factors, which contain a second-sound peak next to the plasmon doublet; the Coulomb drag resistivity, where intrawire equilibration acts as a bottleneck and turns the hydrodynamic drag into an activated temperature dependence; and the near-field heat transfer between the wires. Mode frequencies, spectral weights, linewidths, and the crossover scales of the transport coefficients are expressed through the thermodynamic properties of the liquid and a few dissipative coefficients, providing spectroscopic and transport probes of two-fluid behavior in one-dimensional quantum systems.
\end{abstract}

\maketitle

\section{Introduction}\label{sec:intro}

The theory of hydrodynamic fluctuations was formulated by Landau and Lifshitz
\cite{LandauLifshitz1957,LL9}. Its physical content is based on the fluctuation-dissipation theorem (FDT) \cite{CallenWelton1951}. 
On the scales where the liquid obeys the equations of
hydrodynamics, the same dissipative coefficients that damp the macroscopic flow must also generate thermal fluctuations of the currents. 
Landau and Lifshitz implemented this by adding random Langevin fluxes to the constitutive relations, with
Gaussian white-noise correlators fixed by the viscosities and the thermal conductivity. The corresponding
generalization to superfluids, in the framework of the two-fluid description of $^4$He, was
constructed by Khalatnikov \cite{Khalatnikov1957,KhalatnikovBook}. The construction follows the same principles but is richer. A superfluid has several
independent viscosity coefficients, and the random stresses are cross-correlated with the random
force acting on the superfluid velocity, with all coefficients fixed by Onsager reciprocity \cite{Onsager1931}.

A two-fluid description is not exclusive to superfluids; it arises naturally in normal liquids
whenever a subset of excitations decouples from the rest by virtue of a long mean free path. The
classic example is due to Andreev \cite{Andreev1971}. At temperatures well below the Debye
temperature the wavelength of a thermal phonon in a liquid greatly exceeds the interatomic
distance, and such phonons are weakly damped elementary excitations. The liquid then consists of
two weakly coupled subsystems: the gas of long-wavelength phonons and the remainder of the
liquid, each in local equilibrium with its own temperature and drift velocity on time scales
shorter than the mutual relaxation time. Developing the corresponding two-fluid description,
Andreev predicted qualitatively new phenomena, among them a thermomechanical effect in an ordinary
(nonsuperfluid) liquid and slowly decaying two-liquid modes. In this example a two-fluid description applies
in the frequency window opened by the separation between fast and slow relaxation processes. In a Luttinger liquid, as we recall shortly, precisely such a separation is provided by
the exponential slowness of backscattering.

For electron liquids in solids, the hydrodynamic description in the Fermi-liquid regime goes back
to Gurzhi \cite{Gurzhi1968,GurzhiKopeliovich}. It applies in the collision-dominated limit
$\omega\tau\ll1$, where $\tau$ is the electron-electron collision time responsible for local
equilibration of the liquid. Among its most striking consequences, Gurzhi demonstrated that in
finite-size geometries the resistance is governed by the electron viscosity, so that transport
measurements become probes of the kinetic coefficients of the electron fluid. More recently this
theory was generalized to strongly correlated electron fluids in high-mobility semiconductor
heterostructures, where it was shown that the hydrodynamic resistivity depends also on the thermal
conductivity of the liquid \cite{AKS2011}. In the opposite, quantum limit the hydrodynamic
approach to one-dimensional (1D) electrons was developed primarily in the context of Luttinger
liquids \cite{Haldane1981}, applicable in particular to electrons confined to the lowest subband
of a quantum wire. The quantum hydrodynamics of the Luttinger liquid is special: in the Luttinger
model limit of a strictly linear electron spectrum, the system of interacting fermions maps onto
noninteracting bosons, which physically represent propagating density fluctuations. Precisely
because the model is integrable in this limit, it captures no relaxation: the Luttinger model by
itself possesses no kinetics. This gap was understood and filled over the past decade through the
study of the crossover between the Luttinger-liquid and Fermi-liquid regimes of 1D conductors
\cite{ISG2012,LevchenkoMicklitz2021}. The analysis of relaxation of 1D fermions shows that, even
when the band curvature is retained, two-particle collisions are kinematically ineffective, and
equilibration is controlled by three-particle processes 
\cite{Lunde2007,Micklitz2010,Levchenko2011,Levchenko2011,Teber2012,Ristivojevic2013,Protopopov2014,Rieder2014,Protopopov2015,Ristivojevic2021}. 
These split into two classes: processes
that backscatter a particle and thereby change the numbers of right and left movers, which we
refer to as equilibration, and processes that redistribute momentum and energy at fixed
numbers of movers, which we refer to as thermalization. Equilibration is exponentially slow
at low temperature, since it requires an electron to diffuse through the states near the bottom of
the band, which are statistically inaccessible due to Pauli blocking; its rate carries the
activation factor $e^{-D/T}$, with $D$ of the order of the Fermi energy. Thermalization, in
contrast, is only power-law suppressed in temperature and proceeds at a parametrically faster
rate.

The theory of hydrodynamic fluctuations in two-dimensional electron liquids, including its
generalization to systems lacking Galilean invariance such as graphene and non-Fermi liquids, was developed in
Refs.~\cite{ALA2014,Patel2017,LLA2022,ZL2023,Levchenko2026}, where it was applied to Coulomb drag, plasmon-assisted
transport, and fluctuation-driven heat transfer in double layers. No counterpart of this theory
has been available for Luttinger liquids, and constructing it is the primary purpose of the
present work. It was shown recently that the hydrodynamics of a Luttinger liquid at low temperature is
itself of two-fluid character and closely resembles the two-fluid hydrodynamics of superfluid
$^4$He \cite{MA2019,PSMG2021}. The reason is the separation of time scales described above. On
time scales shorter than the equilibration time, the difference between the numbers of right and
left movers is approximately conserved. The associated zero mode of the liquid plays the role of
the superfluid component, and the gas of thermal excitations forms the normal component.
This is the same structural mechanism as in Andreev's two-liquid description of a normal liquid:
a conservation law that is weakly violated opens a hydrodynamic window with an enlarged set of
slow variables.

The two-fluid theory of 1D quantum liquids was formulated in Ref. \cite{MA2019},
where the coupled hydrodynamic equations for the particle density, energy, momentum, and the
zero-mode velocity of a Luttinger liquid without assuming Galilean invariance were derived. In the window of
frequencies between the slow equilibration rate and the fast thermalization rate the theory
supports two sound modes, the analogs of first and second sound, whose hybridization and
attenuation were also analyzed in detail \cite{MA2017,MA2018,MAhybrid2018}. At frequencies below the
equilibration rate the theory reduces to single-fluid hydrodynamics with anomalous dissipative
coefficients proportional to the exponentially long backscattering time, among them the thermal
conductivity \cite{MR2015}. A parallel development was presented in Refs. \cite{PSMG2021,SPMG2019,GPSM2023}, 
where the same hierarchy of hydrodynamic regimes was derived from the kinetics of fermionic 
quasiparticles, and the anomalous transport emerging at the longest length scales was analyzed within the framework of nonlinear fluctuating hydrodynamics.
The two frameworks describe the same sequence of regimes in different variables, and we make the correspondence
explicit below.

Building on these developments, in this paper we construct the theory of hydrodynamic fluctuations
in the two-fluid regime of a Luttinger liquid. The construction uses only Onsager reciprocity and the fluctuation-dissipation theorem
applied to the entropy production of the two-fluid equations. It yields the full set of Langevin
fluxes, including a mutual-friction noise tied to the backscattering rate, which has no analog in
superfluid helium. We then apply the theory to Coulomb-coupled quantum wires. The paper is organized as follows. In
Sec.~\ref{sec:theory} we review the general Landau-Lifshitz-Khalatnikov construction, formulate
the stochastic two-fluid equations of the Luttinger liquid, and derive their reduction to
single-fluid hydrodynamics with anomalous transport coefficients, for which we provide the
corresponding Langevin fluxes. In Sec.~\ref{sec:applications} we present three applications: the
dynamic structure factors of coupled wires (Sec.~\ref{sec:dsf}), the Coulomb drag resistivity
(Sec.~\ref{sec:drag}), and the near-field heat transfer between wires (Sec.~\ref{sec:nfht}).
Section~\ref{sec:summary} summarizes the results, delimits the validity of the linearized
treatment in light of fluctuation renormalizations of low-dimensional hydrodynamics, and outlines
open problems. Technical details of the fluctuational drag calculations are collected in the
Appendixes.

\section{Stochastic two-fluid hydrodynamics of a Luttinger liquid}
\label{sec:theory}

\subsection{The Landau-Lifshitz-Khalatnikov construction}
\label{sec:LLK}

The construction of fluctuating hydrodynamics is algorithmic once the entropy production of the
dissipative equations is known \cite{LandauLifshitz1957,LL9}. Let the dissipative fluxes $Y_a$ be
linear in the thermodynamic forces $X_a$,
\begin{equation}
Y_a=-\sum_b\gamma_{ab}X_b+y_a,\qquad
T\dot S=\sum_{ab}\gamma_{ab}X_aX_b\geq0,
\label{eq:fluxforce}
\end{equation}
with a symmetric matrix of kinetic coefficients, $\gamma_{ab}=\gamma_{ba}$, as required by Onsager
reciprocity. The fluctuation-dissipation theorem then fixes the Langevin sources $y_a$ to be
Gaussian and white, with correlation functions
\begin{equation}
\langle y_a(x,t)\,y_b(x',t')\rangle=2T\gamma_{ab}\,\delta(x-x')\,\delta(t-t').
\label{eq:LLK}
\end{equation}
For a normal fluid this rule reproduces the Landau-Lifshitz random stresses and heat fluxes. For
superfluid $^4$He, the dissipative two-fluid hydrodynamics involves the shear viscosity $\eta$ of
the normal component, four second-viscosity coefficients $\zeta_{1\ldots4}$ associated with the
relative motion of the components, and the thermal conductivity $\kappa$
\cite{KhalatnikovBook}. In three-dimensional notation the dissipative fluxes read
\begin{align}
\tau_{ik}&=-\eta\Big(\partial_iv_{nk}+\partial_kv_{ni}-\tfrac{2}{3}\delta_{ik}\,{\rm div}\,\bm v_n\Big)
\nonumber\\
&\quad-\delta_{ik}\big[\zeta_1\,{\rm div}(\bm j-\rho\bm v_n)+\zeta_2\,{\rm div}\,\bm v_n\big]+\varsigma_{ik},
\nonumber\\
h&=-\zeta_3\,{\rm div}(\bm j-\rho\bm v_n)-\zeta_4\,{\rm div}\,\bm v_n+\xi,
\nonumber\\
\bm q&=-\kappa\nabla T+\bm g,
\label{eq:helium}
\end{align}
where $\tau_{ik}$ is the dissipative momentum flux, $\bm q$ the heat flux, and the scalar $h$
enters the Josephson equation for the superfluid velocity,
$\partial_t\bm v_s+\nabla(\mu+\tfrac12v_s^2+h)=0$; Onsager symmetry imposes $\zeta_1=\zeta_4$ and
positivity of the entropy production requires $\zeta_1^2\le\zeta_2\zeta_3$. Khalatnikov
\cite{Khalatnikov1957} applied the rule \eqref{eq:LLK} to this set of fluxes and obtained the
noise algebra of the superfluid,
\begin{align}
\langle\varsigma_{ik}\varsigma_{lm}\rangle&=
2T\Big[\eta\big(\delta_{il}\delta_{km}+\delta_{im}\delta_{kl}-\tfrac23\delta_{ik}\delta_{lm}\big)
+\zeta_2\delta_{ik}\delta_{lm}\Big]\delta\delta,
\nonumber\\
\langle\varsigma_{ik}\,\xi\rangle&=2T\zeta_1\delta_{ik}\,\delta\delta,\qquad
\langle\xi\,\xi\rangle=2T\zeta_3\,\delta\delta,
\nonumber\\
\langle g_ig_k\rangle&=2\kappa T^{2}\delta_{ik}\,\delta\delta,
\label{eq:khalatnikov}
\end{align}
where $\delta\delta\equiv\delta(\bm r-\bm r')\delta(t-t')$ and all other cross-correlators vanish
by symmetry. The hallmark of the superfluid case is the cross-correlation between the random
stress and the random force in the Josephson equation, weighted by $\zeta_1=\zeta_4$.
Remarkably, Khalatnikov's paper also introduced a random flux for the current of particles
dissolved in the superfluid (the $^3$He impurity current), with variance proportional to the
diffusion coefficient. This random particle flux is the direct ancestor of the intrinsic-current
noise that we employ for non-Galilean-invariant electron liquids in Sec.~\ref{sec:singlefluid} and
Appendix~\ref{app:sigma}.

\subsection{Langevin extension of the two-fluid theory of Luttinger liquids}
\label{sec:MAstochastic}

The two-fluid hydrodynamics of a 1D quantum liquid \cite{MA2019} differs from that of helium in
three structural respects. First, in one
dimension there is no shear flow, and all dissipative fluxes are scalars or one-dimensional
vectors. Second, Galilean invariance is generically absent for electrons in a solid, so the
particle current is not tied to the momentum density; the two-fluid densities entering the current
and the momentum are independent thermodynamic parameters, and additional dissipative response
appears in the particle and energy currents. Third, the superfluid degree of freedom of the
Luttinger liquid is not topologically protected: backscattering relaxes the difference between the
numbers of right and left movers at the exponentially small rate $\tau_J^{-1}\propto e^{-D/T}$, a
relaxation channel with no counterpart in the hydrodynamics of helium, which must therefore carry
its own Langevin partner.

We follow the notation of Ref.~\cite{MA2019}. The zero mode is characterized by the momentum per
particle $\chi=\pi\hbar J/L$, where $J$ is the winding number of the boson field and $L$ the
system size; the corresponding zero-mode velocity is $\chi/m_0$ with $m_0=\pi\hbar n/vK$, where
$n$ is the particle density, $v$ the excitation velocity, and $K$ the Luttinger parameter. The
gas of elementary excitations moves with its own velocity $u$ and has the mass density
$\rho_{\rm ex}=\pi T^{2}/3\hbar v^{3}$; the momentum density of the liquid is
$p=n\chi+\rho_{\rm ex}(u-\chi/m^{*})$ with $m^{*}=\pi\hbar/\partial_n(vK)$. In the Galilean limit
$m^{*}=m_0=m$. The ideal two-fluid dynamics consists of four continuity equations for the
densities of particles, energy, momentum, and $\chi$; the current conjugate to $\chi$ is the
chemical potential supplemented by a thermal correction,
$j_\chi=(m_0v^{2}/n)\,\delta n+(\partial_nv/v)\,\delta\varepsilon$, so that the $\chi$ equation is
the Josephson equation of the problem. Backscattering enters as a relaxation term
\cite{MA2019},
\begin{equation}
\partial_t\chi+\partial_xj_\chi
=-\frac{1}{\tau_J}\frac{\rho_{\rm ex}}{m_0n}\big(\chi-m_0u\big)
\equiv-\frac{\rho_{\rm ex}}{n\,\tau_J}\,w,
\label{eq:MArelax}
\end{equation}
where $w=\chi/m_0-u$ is the velocity of the zero mode relative to the excitation gas, the
one-dimensional counterpart of the counterflow variable
$\bm j-\rho\bm v_n=\rho_s(\bm v_s-\bm v_n)$ that enters the dissipative fluxes
\eqref{eq:helium} of superfluid helium.

To construct the Langevin extension we consider the free energy of the relative motion.
The energy density of the liquid is
$\mathcal E=n\chi^{2}/2m_0+(\chi/m^{*})p_{\rm ex}+p_{\rm ex}^{2}/2\rho_{\rm ex}+\varepsilon_0(T)$,
where $p_{\rm ex}=\rho_{\rm ex}(u-\chi/m^{*})$ is the momentum of the excitation gas and the
cross term $(\chi/m^{*})p_{\rm ex}$ is the Doppler shift of the excitation energies
$\epsilon_q=v|q|+q\chi/m^{*}$ in the presence of the zero mode \cite{MA2019}. Expanding
$\mathcal E$ at fixed particle density and fixed total momentum $p=n\chi+p_{\rm ex}$ in powers of
$w$, one finds that the minimum lies at $w=0$, to leading order in $\rho_{\rm ex}/nm_0$, and that
\begin{equation}
\delta F=\frac{\rho_{\rm ex}}{2}\,w^{2},
\end{equation}
the kinetic energy of counterflow with the normal mass density $\rho_{\rm ex}$, as in the
superfluid analogy. The relaxation \eqref{eq:MArelax} is therefore a dissipative flux-force pair
in the sense of Eq.~\eqref{eq:fluxforce}, with the mutual-friction force density
$F_{\rm mf}=-\Gamma_Jw+\eta$ acting on the zero-mode component [so that
$n\,\partial_t\chi|_{\rm relax}=F_{\rm mf}$, in accord with Eq.~\eqref{eq:MArelax}], its negative
acting on the excitation gas, and
\begin{equation}
\Gamma_J=\frac{\rho_{\rm ex}}{\tau_J},\qquad
\big\langle\eta(x,t)\,\eta(x',t')\big\rangle=2T\,\Gamma_J\,\delta(x-x')\,\delta(t-t').
\label{eq:GammaJ}
\end{equation}
The force $\pm(-\Gamma_Jw+\eta)$ enters the momentum balances of the two components with opposite
signs, so that the total momentum is conserved microscopically.

More generally, the dissipative sector of the 1D two-fluid theory consistent with spatial-parity
and time-reversal symmetry and with Onsager reciprocity is
\begin{align}
\Pi^{\rm diss}&=-\zeta_2\,\partial_xu-\zeta_1\,\partial_xw+\varsigma, \nonumber\\
h&=-\zeta_4\,\partial_xu-\zeta_3\,\partial_xw+\xi,\nonumber\\
q^{\rm diss}&=-\kappa\,\partial_xT+g,\qquad
F_{\rm mf}=-\Gamma_J\,w+\eta,
\label{eq:fluxes1d}
\end{align}
where $\Pi^{\rm diss}$ is the dissipative momentum flux, $h$ enters the Josephson equation, and
$q^{\rm diss}$ is the dissipative heat flux. The entropy production is the quadratic form
\begin{align}
T\dot S=\;&\zeta_2(\partial_xu)^2+2\zeta_1(\partial_xu)(\partial_xw)+\zeta_3(\partial_xw)^2
\nonumber\\
&+\frac{\kappa}{T}(\partial_xT)^2+\Gamma_J\,w^{2}\;\geq\;0,
\end{align}
which requires $\zeta_1=\zeta_4$, $\zeta_1^{2}\le\zeta_2\zeta_3$, and $\kappa,\Gamma_J\ge0$. The
rule \eqref{eq:LLK} then dictates the complete noise algebra of the stochastic two-fluid theory:
\begin{equation}
\begin{aligned}
\langle\varsigma\varsigma\rangle&=2T\zeta_2\,\delta\delta, &
\langle\varsigma\,\xi\rangle&=2T\zeta_1\,\delta\delta, \\
\langle\xi\xi\rangle&=2T\zeta_3\,\delta\delta, &
\langle gg\rangle&=2T^{2}\kappa\,\delta\delta, \\
\langle\eta\eta\rangle&=2T\,\Gamma_J\,\delta\delta, &
\langle\eta\varsigma\rangle&=\langle\eta\xi\rangle=\langle\eta g\rangle=0 .
\end{aligned}
\label{eq:noise1d}
\end{equation}
The vanishing cross-correlators follow from symmetry: $g$ decouples from $\varsigma$ and $\xi$ by
time-reversal signature, while $\eta$ is a force density, odd under spatial inversion, and cannot
correlate with the parity-even scalar noises. Equations \eqref{eq:fluxes1d}--\eqref{eq:noise1d} are the Luttinger-liquid counterpart
of Khalatnikov's construction. The $\langle\varsigma\xi\rangle$ cross-correlation is the analog of
the $\zeta_1{=}\zeta_4$ correlation in helium. The pair $(\Gamma_J,\eta)$ is spesific: it describes
the random momentum exchange between the zero mode and the excitation gas caused by rare
backscattering events. The coefficients $(\zeta_2,\kappa)$
descend from the fast thermalization processes of the excitation gas; in a wire with several
occupied subbands $\zeta_2$ is the bulk viscosity computed in Ref.~\cite{DGM2018}, while $\zeta_1=\zeta_4$ and $\zeta_3$ involve gradients of the counterflow.
Reference~\cite{MA2019} retains only the relaxational channel \eqref{eq:MArelax} in the two-fluid
window, where the gradient coefficients give subleading damping; Eq.~\eqref{eq:GammaJ} is then the
minimal consistent stochastic model.

By fluctuation-dissipation, the mutual-friction channel also fixes the equilibrium noise of the
persistent-current degree of freedom of a wire: integrating the Josephson equation over a wire of
length $\mathcal L$ yields a Lorentzian spectrum of winding-number fluctuations,
$\langle\delta J\,\delta J\rangle_\omega\propto\Gamma_J/(\omega^{2}+\tau_J^{-2})$, an
independently measurable signature of the backscattering rate.

\subsection{Reduction to a single fluid and the anomalous transport coefficients}
\label{sec:singlefluid}

At frequencies below the equilibration rate, $\omega\tau_J\ll1$, the relative velocity $w$ is a
fast variable that follows the slowly varying thermodynamic forces adiabatically. Balancing the
mutual friction against these forces gives \cite{MA2019}
$\delta\chi=\tau_J\big[(m_0\sigma/\rho_{\rm ex})\partial_xT-(1-m_0/m^{*})\partial_x\mu\big]$, where $\sigma$
is the entropy density. Substituting this response, together with the projected noise, into the
particle and energy currents reproduces single-fluid hydrodynamics with dissipative currents
\begin{align}
j_n&=\frac{p}{m_0}-\gamma_{11}\,\partial_x\mu-\gamma_{12}\,\frac{\partial_xT}{T}+\xi_n,
\nonumber\\
j_\varepsilon&=\frac{p}{m_0}\frac{\partial_n(n^{2}v)}{n^{2}v}\,\varepsilon
-\gamma_{21}\,\partial_x\mu-\gamma_{22}\,\frac{\partial_xT}{T}+\xi_\varepsilon,
\label{eq:MAgamma}
\end{align}
with the matrix of kinetic coefficients \cite{MA2019}
\begin{align}
\gamma_{11}&=\tau_J\rho_{\rm ex}\Big(\frac{1}{m^{*}}-\frac{1}{m_0}\Big)^{2},\quad
\gamma_{12}=\gamma_{21}=2\tau_J\varepsilon\Big(\frac{1}{m^{*}}-\frac{1}{m_0}\Big),
\nonumber\\
\gamma_{22}&=2\tau_J\varepsilon v^{2}=\kappa T .
\label{eq:MAcoeffs}
\end{align}
The matrix is of rank one, $\gamma_{ij}=\tau_J\rho_{\rm ex}a_ia_j$ with $a_1=1/m^{*}-1/m_0$ and
$a_2=v^{2}$ (indeed $\gamma_{11}\gamma_{22}=\gamma_{12}^{2}$, using
$\rho_{\rm ex}=2\varepsilon/v^{2}$), the signature of a single slow relaxation mechanism. All the
coefficients \eqref{eq:MAcoeffs} are proportional to $\tau_J\propto e^{D/T}$ and are therefore
exponentially large at low temperature: the low-frequency hydrodynamics of a clean
single-channel quantum liquid dissipates only through the exponentially rare backscattering. In
the Galilean limit only the thermal conductivity survives, retaining its anomalous magnitude
$\kappa=\pi\tau_JTv/3\hbar$ \cite{MR2015}.

The physical identification of these coefficients is as follows.
Equation~\eqref{eq:MAgamma} shows that $\sigma_{\rm 1D}=e^{2}\gamma_{11}$ is the intrinsic
conductivity of the liquid (the dissipative particle current driven by the gradient of the
electrochemical potential), $\gamma_{22}=\kappa T$ its thermal conductivity, and
$\gamma_{12}=\gamma_{21}$ its intrinsic thermoelectric coefficient, the same triad of
coefficients that characterizes non-Galilean-invariant electron liquids in two dimensions
\cite{LLA2022,ZL2023}. Collecting them into the matrix
\begin{equation}
\hat\Upsilon=
\begin{pmatrix}
\sigma_{\rm 1D}/e^{2} & \gamma_{12}/T\\[2pt]
\gamma_{12}/T & \kappa/T
\end{pmatrix},
\end{equation}
the fluctuation-dissipation theorem equips the currents \eqref{eq:MAgamma} with Langevin fluxes
$\vec\xi=(\xi_n,\xi_s)$, with $\xi_s=\xi_\varepsilon/T$ the random entropy current, correlated as
\begin{equation}
\big\langle\vec\xi(x,t)\otimes\vec\xi^{\,\rm T}(x',t')\big\rangle
=2T\,\hat\Upsilon\,\delta(x-x')\,\delta(t-t'),
\label{eq:Upsilonnoise}
\end{equation}
in complete analogy with the two-dimensional construction of Refs.~\cite{LLA2022,ZL2023}. 
Two remarks are in order. First, the derivation is a consistency check of the stochastic
theory: the single Langevin pair \eqref{eq:GammaJ}, carried through the elimination of $w$,
produces the rank-one noise matrix required by Eq.~\eqref{eq:LLK} for the coefficients
\eqref{eq:MAcoeffs}. Second, because the
coefficients originate from a slow internal relaxation, they are strongly dispersive: repeating
the elimination of $w$ at finite frequency yields the Maxwell-Debye form
\begin{equation}
\gamma_{ij}(\omega)=\frac{\gamma_{ij}}{1-i\omega\tau_J},
\label{eq:debye}
\end{equation}
which interpolates between the anomalous static values and the weak dissipative response
$\re\gamma_{ij}(\omega)\simeq\gamma_{ij}/(\omega\tau_J)^{2}\propto1/\tau_J$ at frequencies inside
the two-fluid window. This dispersion protects the theory from spurious conclusions when the static
coefficients are used outside their domain, and it plays a decisive role in the Coulomb drag
problem of Sec.~\ref{sec:drag} and Appendix~\ref{app:sigma}.

\section{Applications to coupled quantum wires}
\label{sec:applications}

Double-layer systems provide experimentally relevant platform for probing electron dynamics and correlations.
Two conductors separated by a barrier that forbids tunneling and couple through the long-range Coulomb
interaction crosstalk only through fluctuations of their charge and energy densities
\cite{Gramila1991,NarozhnyLevchenko2016}. Every interlayer observable (mutual impedance, drag
voltage, transferred heat) is therefore a functional of the correlation functions of hydrodynamic
fluctuations. The strategy of the linear analysis is standard \cite{ForsterBook}. One linearizes
the stochastic equations about equilibrium, solves for the fluctuating fields in terms of the
Langevin fluxes, and reads off the correlation functions from the noise algebra. The poles of the
response give the collective modes; the weights and widths of the resonances measure the
thermodynamic and kinetic coefficients. We follow this path for the two-fluid Luttinger liquid.

Throughout this section we consider two identical parallel wires of width $a$, separated by the
distance $d\gg a$, with the intrawire and interwire Coulomb kernels
\begin{equation}
U(q)=2e^{2}\ln\frac{1}{|aq|},\qquad
U_{12}(q)=U(q)\,e^{-|q|d},
\label{eq:kernels}
\end{equation}
valid with logarithmic accuracy for $q\ll1/a$. The analysis is controlled by the following
hierarchy of energy scales, evaluated at the interwire momentum $q\sim1/d$ that dominates all
cross-wire phenomena: the fast thermalization rate $\tau_{\rm ex}^{-1}$ of the excitation
gas (a power law in $T$); the exponentially slow equilibration rate
$\tau_J^{-1}\propto e^{-D/T}$; the plasmon frequency $\omega_p(q)=q\sqrt{n^{2}U(q)/\rho}$, with
$\rho=m_0n$; and the second-sound frequency $vq$. The hydrodynamic description of the fluctuation
modes requires $\omega_p(1/d)\,\tau_{\rm ex}\ll1$, i.e., an interwire distance exceeding the
thermalization length. The position of $\tau_J^{-1}$ relative to the mode frequencies then decides
between the two-fluid regime, $\omega\tau_J\gg1$, where the zero mode is dynamically decoupled,
and the single-fluid regime, $\omega\tau_J\ll1$, governed by the anomalous coefficients
\eqref{eq:MAcoeffs}.

\subsection{Dynamic structure factors}
\label{sec:dsf}

\subsubsection{Single-fluid spectra}

\begin{figure*}[t]
\includegraphics[width=\linewidth]{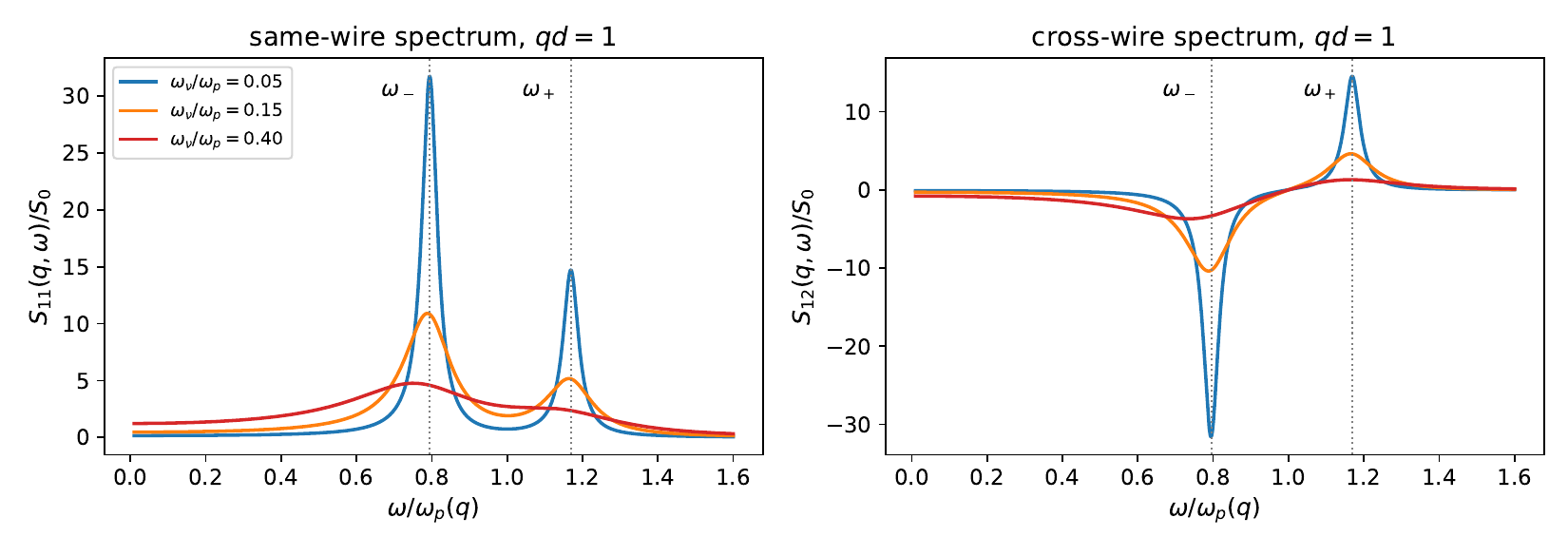}
\caption{Equilibrium density-noise spectra of two coupled wires in the single-fluid regime at
$qd=1$, for three damping ratios $\omega_\nu/\omega_p$, in units of $\omega_p(q)$ and
$S_0=Tn^{2}q^{2}/\rho\,\omega_p^{3}(q)$. Left: same-wire spectrum $S_{11}$, Eq.~\eqref{eq:S12},
showing the plasmon doublet $\omega_\mp$. Right: cross-wire spectrum $S_{12}$: the out-of-phase
plasmon produces anticorrelation, the in-phase mode positive correlation, with total weights fixed
by the sum rules \eqref{eq:sumrule}.}
\label{fig:S1}
\end{figure*}

Consider first the minimal single-fluid model in which each wire is described by the linearized
continuity and Navier--Stokes equations, $\partial_t\delta n+n\partial_x\delta v=0$ and
$\rho\,\partial_t\delta v=-n\partial_x e\delta\Phi+\zeta\partial_x^{2}\delta v+\partial_x\Xi$,
where $\zeta$ is the bulk viscosity of the single-fluid description [it equals $\zeta_2$ of
Eq.~\eqref{eq:fluxes1d} to leading order after the elimination of $w$; in a multichannel wire it
is the fast coefficient \eqref{eq:DGM}], $\Xi$ is its Langevin partner,
$\langle\Xi\Xi\rangle=2\zeta T\delta\delta$, $e\delta\Phi$ is the electrostatic potential, and
the restoring force is dominated by the Coulomb interaction \eqref{eq:kernels}. Symmetric and antisymmetric combinations of the two wires'
densities, $\delta n_\pm=\delta n_1\pm\delta n_2$, decouple, with mode frequencies
$\omega_\pm^{2}=\omega_p^{2}(q)\big(1\pm e^{-|q|d}\big)$ and propagators
\begin{equation}
\Gamma_\pm(q,\omega)=\omega^{2}+i\omega\omega_\nu-\omega_\pm^{2},\qquad
\omega_\nu=\frac{\zeta q^{2}}{\rho}.
\label{eq:Gammapm}
\end{equation}
Solving the stochastic equations and averaging with the noise variances yields the equilibrium
spectra in closed form,
\begin{align}
S_{11}(q,\omega)&=T\zeta\,\frac{n^{2}q^{4}}{\rho^{2}}
\left[\frac{1}{|\Gamma_+|^{2}}+\frac{1}{|\Gamma_-|^{2}}\right],
\nonumber\\
S_{12}(q,\omega)&=T\zeta\,\frac{n^{2}q^{4}}{\rho^{2}}
\left[\frac{1}{|\Gamma_+|^{2}}-\frac{1}{|\Gamma_-|^{2}}\right]
\nonumber\\
&=T\zeta\,\frac{n^{2}q^{4}}{\rho^{2}}\,
\frac{(\omega_+^{2}-\omega_-^{2})(2\omega^{2}-\omega_+^{2}-\omega_-^{2})}{|\Gamma_+|^{2}|\Gamma_-|^{2}},
\label{eq:S12}
\end{align}
with $|\Gamma_\pm|^{2}=(\omega^{2}-\omega_\pm^{2})^{2}+\omega^{2}\omega_\nu^{2}$. The spectra obey
the frequency sum rules
\begin{equation}
\int\frac{\dd\omega}{2\pi}\,S_{11(12)}(q,\omega)
=\frac{Tn^{2}q^{2}}{2\rho}\left(\frac{1}{\omega_+^{2}}\pm\frac{1}{\omega_-^{2}}\right),
\label{eq:sumrule}
\end{equation}
which fix the total spectral weight by the static (screening) correlations and provide an
internal check of the Langevin normalization. Figure~\ref{fig:S1} shows the two spectra.
The same-wire response consists of the plasmon doublet $\omega_\mp(q)$; the cross-wire spectrum
is negative at low frequencies and around the out-of-phase mode $\omega_-$ (the wires
anticorrelate, as screening requires) and positively correlated at the in-phase mode $\omega_+$.
The cross-wire noise spectrum is accessible capacitively, by coupling matched charge sensors or
resonators to segments of the two wires, and requires no current drive.

\subsubsection{Two-fluid spectra}

In the two-fluid window the same observables are computed from the stochastic equations of
Sec.~\ref{sec:MAstochastic}. It is convenient to use as variables the particle density, the
entropy density $\sigma=2\varepsilon/T$, the momentum of the excitation gas
$p_{\rm ex}=\rho_{\rm ex}(u-\chi/m^{*})$, and the zero-mode momentum $\chi$. Per wire, the
linearized system for $x=(\delta n,\delta\sigma,\delta p_{\rm ex},\delta\chi)$ reads
\begin{align}
\partial_t\delta n&=-\partial_x\Big[\tfrac{n}{m_0}\delta\chi+\tfrac{1}{m^{*}}\delta p_{\rm ex}\Big],
\nonumber\\
\partial_t\delta\sigma&=-\partial_x\Big[\sigma u-\tfrac{\kappa_{\rm ex}}{T}\partial_x\delta T+\tfrac{g}{T}\Big],
\nonumber\\
\partial_t\delta p_{\rm ex}&=-\partial_x\Big[\sigma\,\delta T-\zeta_2\partial_xu+\varsigma\Big]
+\Gamma_Jw-\eta,
\nonumber\\
\partial_t\delta\chi&=-\partial_x\Big[\tfrac{m_0v^{2}}{n}\delta n+\delta\mu_{\rm th}
+e\delta\Phi\Big]-\frac{\Gamma_J}{n}w+\frac{\eta}{n},
\label{eq:MA4x4}
\end{align}
where $u=\chi/m^{*}+p_{\rm ex}/\rho_{\rm ex}$, $w=\chi/m_0-u$, and the fluctuations of the
temperature and of the thermal part of the chemical potential follow from the equation of state
of the excitation gas,
$\delta T=(T/\sigma)\,\delta\sigma+T(\partial_nv/v)\,\delta n$ and
$\delta\mu_{\rm th}=T(\partial_nv/v)\,\delta\sigma+\varepsilon(\partial_n^{2}v/v)\,\delta n$.
In these variables the thermodynamic consistency of the system is manifest: by the Gibbs-Duhem
relation of the excitation subsystem, the pressure and chemical-potential gradients acting on the
excitation gas combine into the single term $\sigma\partial_x\delta T$. In the variables
$(\delta n,\delta\varepsilon,\delta p,\delta\chi)$ of Ref.~\cite{MA2019} the same equations
contain the density derivatives of the thermal pressure and of $\mu_{\rm th}$, which are formally
of order $\varepsilon$ but cannot be dropped: after the cancellation $\delta p-n\,\delta\chi$ they
act as $O(1)$ forces on the light excitation gas, and the equilibrium sum rules hold only when
they are kept. Here $\kappa_{\rm ex}$ and $\zeta_2$ are the thermal conductivity and the
bulk viscosity of the excitation gas (fast, power-law coefficients, to be distinguished from the
anomalous $\kappa$ of Eq.~\eqref{eq:MAcoeffs}), $e\delta\Phi_i=U(q)[\delta n_i+e^{-qd}\delta n_j]$
is the electrostatic potential in wire $i$ produced by the charges of both wires, and the noises
obey Eq.~\eqref{eq:noise1d} with $\kappa\to\kappa_{\rm ex}$. Writing the resulting $8\times8$
linear system as $M(q,\omega)\,x=B\,\xi$, with $\xi=(\varsigma_1,g_1,\eta_1,\varsigma_2,g_2,\eta_2)$
the Langevin sources and $\mathcal G=M^{-1}B$, the dynamic structure factors are
$S_{ij}(q,\omega)=\sum_{a}\mathcal G_{ia}(q,\omega)\Sigma_{aa}\mathcal G^{*}_{ja}(q,\omega)$ with
$\Sigma={\rm diag}(2T\zeta_2,2T^{2}\kappa_{\rm ex},2T\Gamma_J)^{\oplus2}$. Near each pole pair the
exact expression collapses to a Lorentzian. It can be verified that the spectra obey the equilibrium sum rules
$\int\frac{\dd\omega}{2\pi}S_{ij}=T\,[(m_0v^{2}/n)\hat1+\hat U]^{-1}_{ij}$, with
$\hat U_{ij}=U(q)e^{-|q|d(1-\delta_{ij})}$, which reduce to Eq.~\eqref{eq:sumrule} when the
intrinsic pressure $m_0v^{2}/n$ is neglected against $U$.

\begin{figure*}[t]
\includegraphics[width=\linewidth]{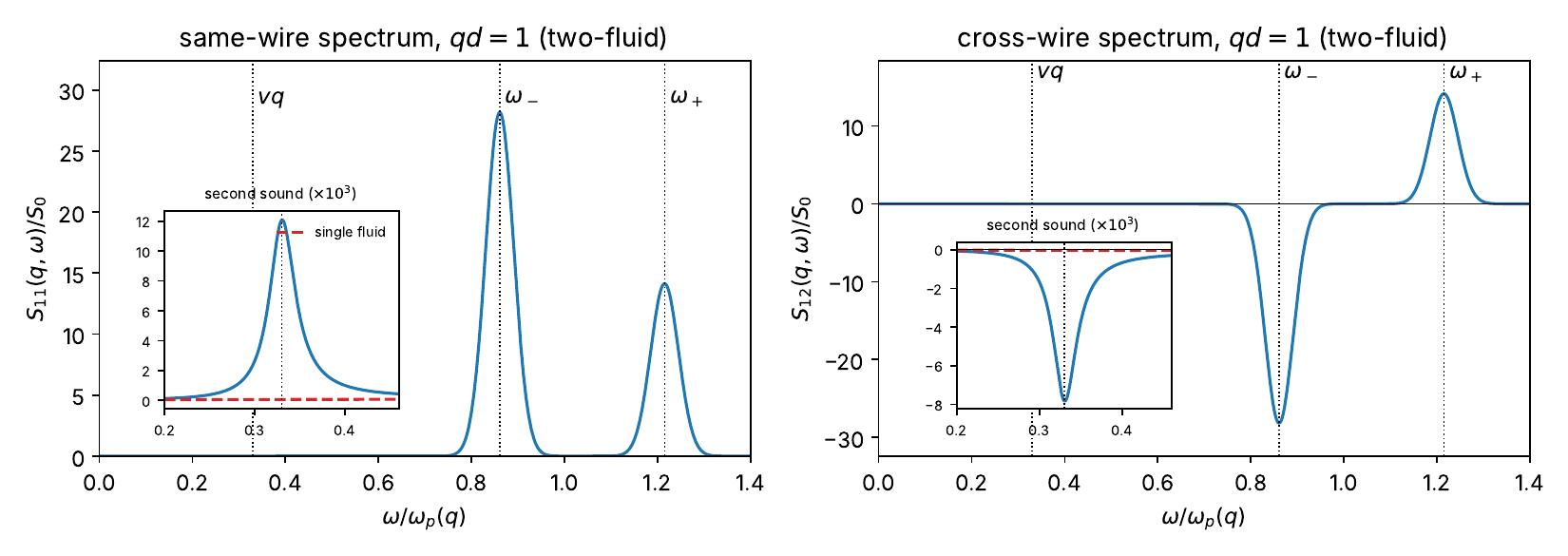}
\caption{Two-fluid density spectra of coupled wires at $qd=1$, computed from the stochastic
system \eqref{eq:MA4x4}. Frequencies are in units of the plasmon frequency $\omega_p(q)$ and
spectra in units of $S_0=Tn^{2}q^{2}/\rho\,\omega_p^{3}(q)$; parameters: $nd=5$,
$T/\omega_p(1/d)=0.10$, $vq/\omega_p(q)=0.33$, $K=0.7$, $m^{*}/m_0=1.4$, $n\partial_nv/v=0.5$,
$\tau_Jv/d=200$, and fast dissipation set by the second-sound damping ratios
$\zeta_2q^{2}/\rho_{\rm ex}vq=\kappa_{\rm ex}q^{2}/c_vvq=0.05$. The plasmon lines are narrower
than the second-sound line by a factor of order $\rho_{\rm ex}/\rho$; the main panels therefore
show the spectra at a finite frequency resolution $\delta\omega=0.03\,\omega_p$ (Gaussian), as in
a bandwidth-limited noise measurement, which brings the plasmon peaks to the scale of
Fig.~\ref{fig:S1}. Left: same-wire spectrum; the plasmon doublet $\omega_\mp$ is accompanied by
the second-sound resonance at $\omega=vq$, shown magnified in the inset together
with the single-fluid spectrum (dashed), which has no such feature. Right: cross-wire spectrum;
the second-sound feature appears in the anticorrelated phase (inset).}
\label{fig:S2}
\end{figure*}

Figure~\ref{fig:S2} shows the result, and three features deserve emphasis. First, there are now
four resonances per parity sector: the long-range Coulomb interaction pushes the charge
modes up to the plasmon doublet $\omega_\pm(q)\gg vq$, while the second sound (ballistic heat
propagation, which for a single excitation branch travels at the excitation velocity itself,
$u_2=v$ \cite{MA2017}) remains far below. In a charged wire the two sounds are thus
parametrically separated, in contrast to neutral 1D liquids where $u_1\simeq u_2\simeq v$ and the
modes hybridize strongly \cite{MAhybrid2018}. Second, the weight of the second-sound resonance in the
density channel is controlled by the thermal-pressure couplings $\partial_n(nv)/v$ and
$\partial_nv/v$ and by the non-Galilean parameter $(1/m^{*}-1/m_0)$; numerically it scales as the
square of these couplings, and a reduced weight survives even at the Galilean point through the
temperature dependence of the excitation-gas pressure. For the parameters of Fig.~\ref{fig:S2}
the second-sound peak in $S_{11}$ carries $\sim2\times10^{-4}$ of the plasmon weight: small but
visible, and its position measures the excitation velocity directly while its width measures the
fast-process coefficients. The plasmon lines, in turn, are much sharper than the second-sound
line: in the two-fluid window all dissipation originates in the light excitation gas, so the
plasmon damping is smaller than the second-sound damping by a factor of order
$\rho_{\rm ex}/\rho$. This is why the main panels of Fig.~\ref{fig:S2} are shown with a finite
frequency resolution, while the insets show the unbroadened second-sound line. Third, the cross-wire spectrum retains the
anticorrelated/correlated plasmon structure of Eq.~\eqref{eq:S12} and acquires the second-sound
feature in the screening-anticorrelated phase; its detection in the cross-wire noise would be a
direct, drive-free signature of two-fluid behavior of a Luttinger liquid.

The linewidths in $S_{11}$ and $S_{12}$ are fixed by the coefficient set
$\{\zeta_2,\zeta_3,\zeta_1{=}\zeta_4,\kappa,\Gamma_J\}$. Measuring the attenuation of first and
second sound (in wires through the noise spectra above, or in cold-atom realizations of 1D gases
through Bragg spectroscopy) therefore amounts to a direct measurement of the kinetic coefficients
of the two-fluid theory, with the multichannel bulk viscosity of Ref.~\cite{DGM2018} controlling
the first-sound width in few-channel wires. In this sense the fluctuation spectra realize a
coefficient spectroscopy of 1D hydrodynamics that does not rely on any transport
measurement.

It is instructive to place these results against the theory of the coherent dynamic response of
1D fermions initiated in Ref. \cite{PKKG2006} and completed
within the nonlinear Luttinger-liquid program \cite{ISG2012}. In that theory (zero temperature,
no equilibrating/thermalizing collisions) band curvature spreads the delta-function peak of the linear Luttinger model into
a band of width $\delta\omega\simeq q^{2}/m^{*}$ between kinematic edges, and interactions convert
the edges into power-law singularities with interaction-dependent exponents, a 1D counterpart of
the x-ray edge problem. The spectra computed here are the collision-dominated counterpart of the
same observable: they apply when the thermal collision rate exceeds the curvature dispersion,
\begin{equation}
\tau_{\rm ex}^{-1}(T)\gg\frac{q^{2}}{m^{*}}
\quad\Longleftrightarrow\quad
q\ll q^{*}(T)=\sqrt{\frac{m^{*}}{\tau_{\rm ex}(T)}},
\label{eq:qstar}
\end{equation}
i.e., precisely when the quasiparticle interference that builds the edge singularities is
scrambled within one period of the mode. The edge structure then heals into the Lorentzian
resonances of Fig.~\ref{fig:S2}: spectral support is bounded by damping rather than kinematics,
edge exponents are replaced by mode widths expressible through transport coefficients, and a
qualitatively new feature (the second-sound satellite, a purely finite-temperature,
collision-enabled mode) appears with no counterpart in the coherent response. 
The two descriptions are complementary limits of one crossover in the $(q,T)$ plane; its
quantitative theory near $q\sim q^{*}$ remains open. The cross-wire spectrum $S_{12}$ has no
analog in the nonlinear Luttinger-liquid theory, since it requires the fluctuation-dissipation
structure of two coupled dissipative liquids.

\subsection{Coulomb drag}
\label{sec:drag}

Coulomb drag is the canonical probe of interlayer correlations \cite{Pogrebinskii1977,Price1983,
Gramila1991,NarozhnyLevchenko2016}: a current $I$ driven through an active wire induces, through
momentum transfer by interwire Coulomb scattering, a voltage in an electrically isolated passive
wire, and the drag resistivity $r_D$ measures the rate of that momentum transfer. In the
hydrodynamic regime of two-dimensional double layers the drag is carried by thermal fluctuations
of the densities and is expressible through the kinetic coefficients of the electron fluid
\cite{ALA2014}. The central question for quantum wires is how this fluctuational mechanism fares
against the two-fluid structure of the 1D liquid.

\subsubsection{Momentum balance and the equilibration bottleneck}

\begin{figure}[t]
\includegraphics[width=\columnwidth]{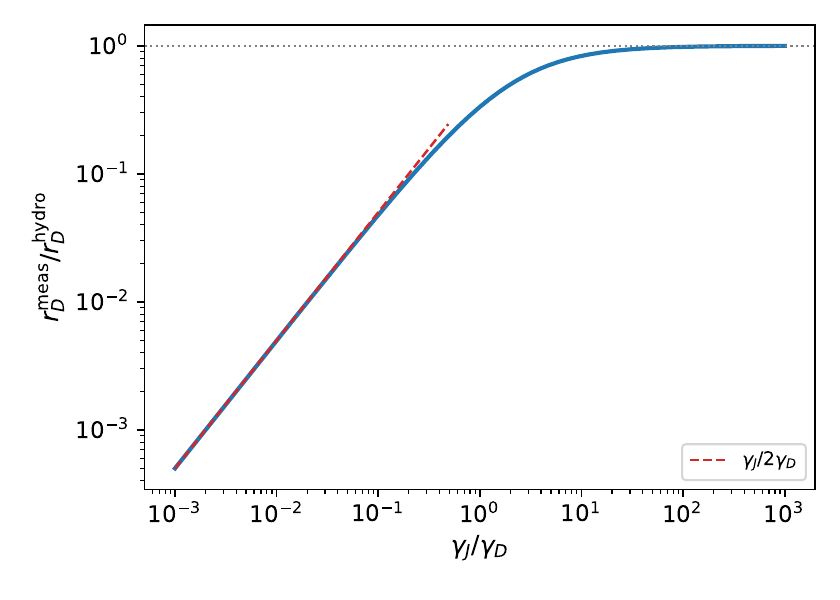}
\caption{Measured dc drag, Eq.~\eqref{eq:series}, versus the ratio of the equilibration rate
$\gamma_J$ to the interwire friction rate $\gamma_D$. The hydrodynamic result forms the plateau;
single-channel Luttinger liquids at low temperature reside deep on the backscattering-limited
tail $r_D^{\rm meas}\simeq\gamma_J/2e^{2}n^{2}\propto e^{-D/T}$.}
\label{fig:drag}
\end{figure}

The measured dc drag follows from a transparent momentum balance on the four momentum reservoirs
of the coupled system: the zero mode and the excitation gas of each wire (details in
Appendix~\ref{app:balance}). A uniform electric field couples to the total charge and therefore
accelerates the zero mode; long-wavelength excitations are charge-neutral density waves, so the
excitation gases are driven only through internal friction. Momentum thus flows along a chain of
three links: from the zero mode of the active wire to its excitation gas at the equilibration rate
$\gamma_J=\rho_{\rm ex}/\tau_J$ [Eq.~\eqref{eq:GammaJ}; in the balance equations of this
section and of Appendix~\ref{app:balance} we write $\gamma_J$ for the mutual-friction coefficient
$\Gamma_J$]; from the excitation gas of wire 1 to that
of wire 2 at the interwire friction rate $\gamma_D=e^{2}n^{2}r_D^{\rm hydro}$, where
$r_D^{\rm hydro}$ is the fluctuational (hydrodynamic) drag computed in
Appendix~\ref{app:singlefluid}; and from the excitation gas of wire 2 into its zero mode, again at
rate $\gamma_J$. In the stationary state the weakest link controls the chain, and the solution of
the linear balance equations gives the exact series form
\begin{equation}
r_D^{\rm meas}=\frac{1}{e^{2}n^{2}}
\left[\frac{1}{\gamma_J}+\frac{1}{\gamma_D}+\frac{1}{\gamma_J}\right]^{-1}
=r_D^{\rm hydro}\,\frac{\gamma_J}{\gamma_J+2\gamma_D}.
\label{eq:series}
\end{equation}
The two limits are shown in Fig.~\ref{fig:drag}. For fast intrawire equilibration,
$\gamma_J\gg\gamma_D$, the hydrodynamic result is recovered in full,
$r_D^{\rm meas}\to r_D^{\rm hydro}$. In the opposite, generic low-temperature limit the measured
drag saturates at
\begin{equation}
r_D^{\rm meas}\simeq\frac{\gamma_J}{2e^{2}n^{2}}
=\frac{\rho_{\rm ex}}{2e^{2}n^{2}\tau_J}\;\propto\;e^{-D/T},
\label{eq:activated}
\end{equation}
set by the backscattering rate alone: the interwire friction $\gamma_D$ drops
out entirely. Physically, the drive current rides in the frictionless zero-mode channel, and the
momentum transfer to the passive wire's charge sector is bottlenecked at both ends of the chain by
the exponentially slow conversion between the zero mode and the excitations, exactly as a
superconducting layer short-circuits drag in a metallic double layer. Equivalently, the state with
a nonzero winding number and both excitation gases at rest is a constrained equilibrium of the
coupled system, since forward-scattering interwire coupling conserves the winding number of each
wire; and constrained equilibria exert no drag. A dc drag voltage requires degrading $J_1$, which
only backscattering can accomplish.

This conclusion agrees with the microscopic theory of drag between clean wires and provides
its hydrodynamic counterpart. In Ref.~\cite{PMGA2003}
it was shown that the forward-scattering (small momentum transfer) drag exists only by
virtue of band curvature, with $r\propto T^{2}$ for identical and $T^{5}$ for strongly different
wires. Subsequently, Ref.~\cite{DGP2012} solved the coupled kinetic equations with three-particle
collisions and showed that these results do not survive as a true dc drag. Without intrawire
right-left equilibration the dc response from forward scattering vanishes; the measured drag is
limited by processes involving cold electrons near the band bottom and becomes activated in
temperature. In one regime it is independent of the interwire coupling altogether, as
Eq.~\eqref{eq:series} predicts for $\gamma_J\ll\gamma_D$. The hydrodynamic formulation adds two
things. It is nonperturbative in the intrawire interaction, and it expresses both sides of the
crossover \eqref{eq:series} through thermodynamic quantities and a few kinetic coefficients,
without reference to a microscopic collision model. The remaining channel, direct
interwire backscattering with momentum transfer $2k_F$, is suppressed as $e^{-2k_Fd}$
\cite{KlesseStern2000,NazarovAverin1998} and is neglected throughout, consistent with the
long-wavelength interwire coupling \eqref{eq:kernels}; its place in the two-fluid picture is
discussed in Sec.~\ref{sec:summary}. For wires of finite length $L$ the same
balance applies with the additional channel of momentum exchange with the leads: for
$L\ll\ell_J=v\tau_J$ the excitation gas is anchored to the stationary contacts and the drag is
suppressed further; since $\ell_J$ is exponentially long, realistic single-channel wires are deep
in this regime. This hierarchy of regimes offers a framework for interpreting the 1D-1D drag experiments in
vertically integrated quantum wires \cite{Laroche2011,Laroche2014,Makaju2025,Zheng2025,Zheng2026},
which are short on the scale of $\ell_J$ and gated into the few-mode regime.

\subsubsection{Window of validity of the hydrodynamic drag}

Collecting the conditions, the fluctuational result $r_D^{\rm meas}\simeq r_D^{\rm hydro}$
describes the measured dc drag if and only if, at the operating temperature: (W1) the fluctuation
modes at the interwire scale are hydrodynamic, $d\gg\ell_{\rm ex}(T)$ and
$\omega_p(1/d)\tau_{\rm ex}\ll1$; (W2) the liquid is a single fluid at the drag-dominant
frequencies, $\omega_p(1/d)\tau_J\ll1$; and (W3) there is no zero-mode bottleneck,
$\gamma_J\gg\gamma_D$, together with $L\gg\ell_J$. Since $\tau_J\propto e^{D/T}$ with $D\sim E_F$
for one channel, conditions (W2) and (W3) fail exponentially at low temperature in a strictly
single-channel wire. They become attainable in wires with several occupied subbands, where the
activation barrier is set by the depth of the shallowest subband, $D\to\mu-\Delta_N$, with
$\Delta_N$ the bottom of the highest occupied subband, and can be
modest. This explains why the hydrodynamic calculation works better for a few-mode wire: 
the key is the rapid equilibration of the zero modes, rather than the finite viscosity itself.

The bulk viscosity entering $r_D^{\rm hydro}$ deserves the same scrutiny. In a multichannel wire,
two-particle collisions between electrons of different subbands provide power-law relaxation and
a finite bulk viscosity \cite{DGM2018},
\begin{equation}
\zeta=\frac{8\tau_{\rm ic}}{h}\sum_i\frac{1}{v_i}
\Bigg(\frac{\sum_j\Delta_j/v_j}{\sum_j1/v_j}-\Delta_i\Bigg)^{\!2}
\xrightarrow[N=2]{}nm\,\tau_{\rm ic}(v_1-v_2)^{2},
\label{eq:DGM}
\end{equation}
where $\Delta_i$ is the energy of the bottom of subband $i$, $v_i=\sqrt{2(\mu-\Delta_i)/m}$
its Fermi velocity, and $\tau_{\rm ic}$ the interchannel collision time, with
$\tau_{\rm ic}^{-1}\propto T$, so that $\zeta\propto1/T$; Eq.~\eqref{eq:DGM} holds for
$T\ll\mu-\Delta_i$ \cite{DGM2018}. This is the fast coefficient
appropriate to the drag kernel. Notably, the interchannel collisions behind Eq.~\eqref{eq:DGM}
conserve the total right-left imbalance: they render $\zeta$ finite but do not by themselves
equilibrate the zero mode. The finiteness of the viscosity and the lifting of the bottleneck are
independent requirements, both of which a few-channel wire can satisfy. In a strictly
single-channel wire, by contrast, all static single-fluid coefficients are the anomalous ones of
Eq.~\eqref{eq:MAcoeffs}, valid only for $\omega\tau_J\ll1$, while the drag kernel is dominated by
$\omega\sim\omega_\pm(1/d)\gg\tau_J^{-1}$: there is no regime in which the single-fluid drag
formula with a meaningful static viscosity applies, and the correct dc answer is the
backscattering-limited Eq.~\eqref{eq:activated}. The same conclusion can be reached within hydrodynamics by keeping the frequency dispersion
\eqref{eq:debye} of the anomalous coefficients: the intrinsic-conductivity channel then reproduces
the activated law of Eq.~\eqref{eq:activated}, with a prefactor that carries the non-Galilean
weight $(m_0/m^{*}-1)^{2}$, as shown in Appendix~\ref{app:sigma}. The hydrodynamic drag itself, $r_D^{\rm hydro}$, receives three additive
fluctuational contributions (viscous, thermal, and intrinsic-current) whose calculation,
regimes, and relative weights are presented in Appendixes~\ref{app:singlefluid}--\ref{app:thermal}.
Here we quote the viscous result for reference:
\begin{equation}
r_D^{(\zeta)}=\frac{1}{\pi^{2}e^{4}}\,\frac{T\zeta/n}{(nd)^{3}}\,
f\!\Big(\beta,\frac{a}{d}\Big),\qquad
\beta=\frac{\zeta}{\rho\,\Omega d^{2}},
\label{eq:rdvisc}
\end{equation}
with $\Omega^{2}=2e^{2}n^{2}/\rho d^{2}$; the scaling function $f$ is given in closed form in
Appendix~\ref{app:singlefluid}, with the limits
$f\simeq(\pi/12)\ln^{3}(\sqrt{\Lambda}/\beta)/\Lambda$ for underdamped plasmons ($\beta\ll1$,
$\Lambda=\ln d/a$), the 1D counterpart of the logarithmic plasmon enhancement in double
layers \cite{ALA2014}, and $f\to\pi\zeta_R(3)/16\Lambda$ in the overdamped limit,
where $\zeta_R$ is the Riemann zeta function.

\subsubsection{Finite-frequency transresistivity}

\begin{figure}[t]
\includegraphics[width=\columnwidth]{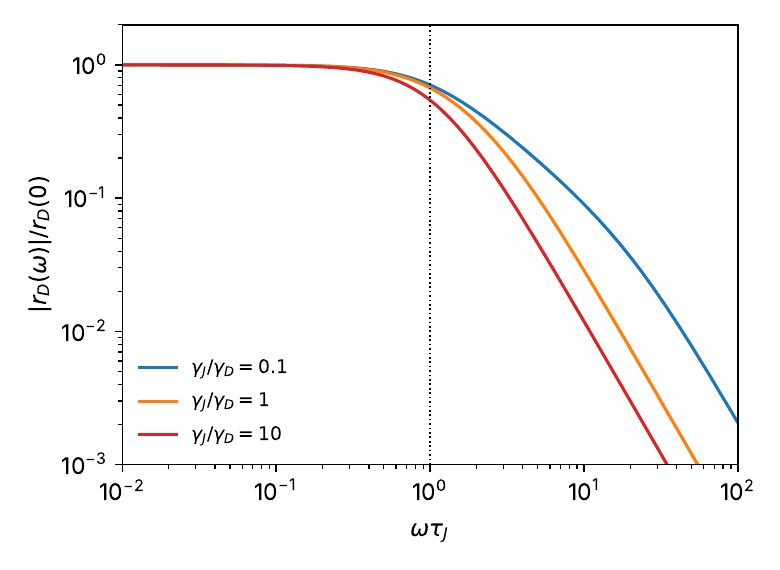}
\caption{Magnitude of the finite-frequency transresistivity, Eq.~\eqref{eq:acdrag}, for
three ratios $\gamma_J/\gamma_D$. The knee at $\omega\tau_J\sim1$ provides direct spectroscopy of the
zero-mode equilibration rate; the plateau height measures the series combination of $\gamma_J$
and $\gamma_D$.}
\label{fig:ac}
\end{figure}

The frequency dependence of the drag provides direct spectroscopy of the equilibration rate.
Retaining the inertia of the excitation gas in the momentum-balance equations
(Appendix~\ref{app:balance}) gives, for a drive of frequency $\omega$,
\begin{equation}
r_D(\omega)=r_D(0)\,
\frac{\gamma_J(\gamma_J+2\gamma_D)}
{\big(\gamma_J-i\omega\rho_{\rm ex}\big)\big(\gamma_J+2\gamma_D-i\omega\rho_{\rm ex}\big)} ,
\label{eq:acdrag}
\end{equation}
whose dissipative part is a product of two Lorentzian factors with knees at
$\omega\sim\tau_J^{-1}$ and $\omega\sim(\gamma_J+2\gamma_D)/\rho_{\rm ex}$
(Fig.~\ref{fig:ac}). The ac response does not recover the bare hydrodynamic value at $\omega\tau_J\gg1$. The
fluctuation ``wind'' that drives the drag must still be established through the $\gamma_J$ links,
and $r_D(\omega)$ decreases beyond the knee. Fig.~\ref{fig:ac} shows the magnitude $|r_D(\omega)|$.
Eq.~\eqref{eq:acdrag} offers a measurement protocol: the knee frequency measures
$\tau_J^{-1}(T)$, with its activated temperature dependence, while the plateau height measures
the series combination of $\gamma_J$ and $\gamma_D$. A measurement of both, confronted with the
theoretical $\gamma_D=e^{2}n^{2}r_D^{\rm hydro}$, would separately test the fluctuational and the
equilibration ingredients of the theory.

\subsection{Near-field heat transfer}
\label{sec:nfht}

The third application concerns the transfer of energy, rather than momentum, between the wires.
Two conductors held at different temperatures exchange heat through the fluctuating Coulomb
fields even in the absence of tunneling and radiation---the near-field analog of radiative heat
transfer \cite{PolderVanHove1971,VolokitinPersson2007,Mahan2017}, studied for two-dimensional double layers
in Refs.~\cite{Kamenev2020,Basko2021,LLA2022}. 
In wires, this observable has a particular advantage. Since heat is carried by the excitation gases at both ends of the transfer chain, 
the near-field heat conductance is not limited by the zero-mode bottleneck. 
This makes it useful indicator of the stochastic theory across densities and temperatures.
 In this precise sense the energy channel plays for wires the
role that charge neutrality plays for two-dimensional layers, where it decouples the observable
from the hydrodynamic zero mode \cite{LLA2022}. The macroscopic observables of a thermal-drag
experiment, induced temperature profiles, dragged heat currents, and thermal locking in long
systems, follow from the heat conductance computed here and from the drag coefficient of
Sec.~\ref{sec:drag} by the force-balance construction of Ref.~\cite{LLA2022}, which carries over
to one-dimensional geometry.

With each wire's Langevin fluxes taken at its own temperature, the interwire energy flux per unit
length is
\begin{equation}
J_E=\frac{e}{2}\big\langle\delta n_2\,\partial_t\delta\phi_2-\delta n_1\,\partial_t\delta\phi_1\big\rangle
=\!\int\!\frac{\dd q\,\dd\omega}{4\pi^{2}}\Big(\!-\frac{i\omega}{2}\Big)U_{12}(q)\,\mathcal D_E,
\label{eq:JE}
\end{equation}
where $\mathcal D_E=\langle\delta n_+\delta n_-\rangle$ is the mixed-parity correlator, nonzero
only for $T_1\neq T_2$ because the symmetric and antisymmetric noise combinations then correlate,
e.g., $\langle\Xi_+\Xi_-\rangle=2\zeta(T_1-T_2)$. Evaluating $\mathcal D_E$ within the
single-fluid theory with both the viscous and the intrinsic-current noise channels yields the
closed formula for the heat conductance per unit length, $\varkappa_{12}=J_E/(T_1-T_2)$:
\begin{align}
\varkappa_{12}&=\int\frac{\dd q\,\dd\omega}{4\pi^{2}}\;U_{12}(q)\,\omega^{2}
\Big[\dfrac{n^{2}q^{4}}{\rho^{2}}\zeta+q^{2}(\omega^{2}+\omega_\nu^{2})\gamma_{11}\Big] \nonumber \\ 
&\times\frac{\Big[\omega_\nu\big(\omega_+^{2}-\omega_-^{2}\big)+(\omega^{2}+\omega_\nu^{2})\big(\gamma_+-\gamma_-\big)\Big]}
{|\tilde\Gamma_+|^{2}\,|\tilde\Gamma_-|^{2}},
\label{eq:kappa12}
\end{align}
where $\gamma_\pm=\gamma_{11}q^{2}U(q)(1\pm e^{-qd})$ are the Maxwell relaxation rates of the
intrinsic-conductivity channel and
$\tilde\Gamma_\pm=\omega^{2}-\omega_\pm^{2}-\omega_\nu\gamma_\pm+i\omega(\omega_\nu+\gamma_\pm)$
generalize Eq.~\eqref{eq:Gammapm}; note that the integrand is positive definite. The kernel
of Eq.~\eqref{eq:kappa12} is the structural twin of the drag kernel derived in
Appendix~\ref{app:sigma}, with one telling replacement, $q^{2}\to\omega^{2}$, in the coupling:
momentum transfer weighs wave vector, energy transfer weighs frequency. In the relaxational
regime, $\gamma_\pm\gg\omega_\pm,\omega_\nu$, the frequency integral gives
$\pi(\gamma_+-\gamma_-)/(\gamma_++\gamma_-)$ per unit noise weight and
\begin{equation}
\varkappa_{12}^{(\sigma)}\simeq\frac{\gamma_{11}}{2\pi}\!\int\!\dd q\,q^{2}U(q)e^{-2qd}
\simeq\frac{\sigma_{\rm 1D}\ln(d/a)}{4\pi\,d^{3}},
\label{eq:kappasigma}
\end{equation}
the direct one-dimensional analog of the result
$\varkappa_{\rm th}=\sigma/8\varepsilon d^{3}$ for graphene double layers at charge neutrality
\cite{LLA2022}: heat conductance proportional to the intrinsic conductivity, whereas the drag of
Appendix~\ref{app:sigma} is inversely proportional to it.

\begin{figure*}[t]
\includegraphics[width=\linewidth]{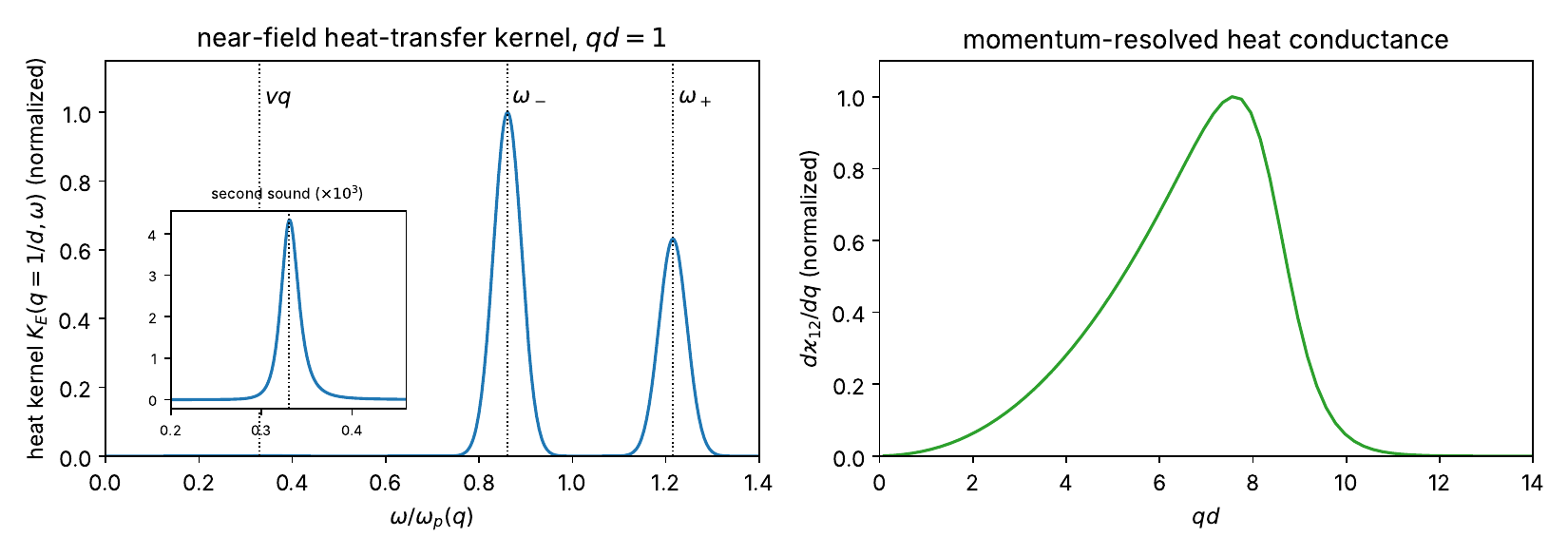}
\caption{Left: near-field heat-transfer kernel $K_E(q,\omega)$ of the stochastic two-fluid
theory at $qd=1$, computed from the two-temperature version of Eq.~\eqref{eq:MA4x4} for the
parameters of Fig.~\ref{fig:S2} and shown with the same frequency resolution
$\delta\omega=0.03\,\omega_p$; the plasmon doublet is accompanied by the second-sound peak at
$\omega=vq$ (inset: intrinsic line shape, magnified). Right: momentum-resolved interwire heat conductance
$d\varkappa_{12}/dq$. With the sharp plasmons of the two-fluid window the integral is dominated
by nearly degenerate plasmon pairs at $qd\gg1$, up to the damping cutoff
$qd\simeq\ln(\sqrt{\Lambda}/\beta_{\rm eff})$ [cf.\ Eq.~\eqref{eq:fasym}], here $qd\simeq10$.}
\label{fig:heat}
\end{figure*}

In the two-fluid window the same energy flux is computed from the stochastic system
\eqref{eq:MA4x4} with per-wire temperatures. The resulting kernel, shown in Fig.~\ref{fig:heat},
exhibits peaks at both plasmons and at the second-sound frequency $\omega=vq$.
For identical wires the second-sound resonances of the two wires are exactly
degenerate, so this channel's strength is set by the square of its (small) density weight divided
by its (small) width; for moderate thermal-pressure couplings the plasmon exchange dominates the
integrated conductance. The bilayer intuition that the
thermal channel dominates \cite{ALA2014,LLA2022} does not carry over automatically to charged
wires: what was a soft diffusive pole in two dimensions is here a propagating mode at finite
frequency, no longer parametrically soft compared to the plasmon in the weight-bearing region
$qd\gtrsim1$. The second sound remains, however, a sharp spectroscopic feature of the heat-transfer
kernel. As for the viscous drag, Eq.~\eqref{eq:fasym}, the integrated conductance is
dominated by nearly degenerate plasmon pairs over the broad interval
$1\lesssim qd\lesssim\ln(\sqrt{\Lambda}/\beta_{\rm eff})$, where $\beta_{\rm eff}$ is the plasmon
damping ratio (Fig.~\ref{fig:heat}, right). The only previous calculation of thermal drag
between Luttinger liquids \cite{Bhandari2018} used the Gaussian (linear) Luttinger model for short
coupled segments. Equation~\eqref{eq:kappa12} reduces to a Gaussian plasmon-exchange result in
the zero-width limit, and extends it with the dissipative and two-fluid structure and with the
equilibration physics that the linear model cannot capture.

\section{Summary, conclusions, and outlook}
\label{sec:summary}

We have constructed the theory of hydrodynamic fluctuations for Luttinger liquids in their
two-fluid regime. The construction is built from the two-fluid equations, Onsager reciprocity, and
the fluctuation-dissipation theorem. Its output is the set of Langevin fluxes,
Eq.~\eqref{eq:noise1d}, which includes the cross-correlated stress and Josephson noises known
from superfluid helium \cite{Khalatnikov1957} and the mutual-friction noise, Eq.~\eqref{eq:GammaJ},
specific to one dimension, where the superfluid degree of freedom relaxes through rare
backscattering. In the single-fluid regime the stochastic theory reproduces the anomalous
transport coefficients of Ref.~\cite{MA2019}, now supplemented by the corresponding Langevin
currents, Eq.~\eqref{eq:Upsilonnoise}, and by their Maxwell-Debye dispersion,
Eq.~\eqref{eq:debye}. Three applications were developed in detail. The dynamic structure factors
of coupled wires display, alongside the plasmon doublet, a second-sound resonance whose position
measures the excitation velocity and whose width measures the fast kinetic coefficients; the
long-range Coulomb interaction separates the two sounds parametrically, making the wire geometry
a useful device for two-fluid spectroscopy. The Coulomb drag between the wires obeys the series law \eqref{eq:series}. The
hydrodynamic drag sets the overall scale, which is exposed only when intrawire equilibration is
faster than the interwire momentum transfer. In the generic low-temperature regime the measured
drag collapses to the activated, backscattering-limited value \eqref{eq:activated}. This agrees
with the kinetic theory of Ref.~\cite{DGP2012}, but is derived here nonperturbatively in the
intrawire interaction and expressed through thermodynamic quantities. The near-field heat
transfer, Eq.~\eqref{eq:kappa12}, is free of the equilibration bottleneck and is the most robust
fluctuational observable of the pair of wires; in the relaxational regime the heat conductance
is proportional to the intrinsic conductivity, Eq.~\eqref{eq:kappasigma}.

Several remarks delimit the validity of the linearized treatment. It is known since the work of
Andreev \cite{Andreev1978,Andreev1980} that the hydrodynamics of low-dimensional liquids is
renormalized by its own thermal fluctuations: in two dimensions the kinetic coefficients acquire
logarithmic scale dependence, and in one dimension sound absorption scales as $\omega^{3/2}$
rather than $\omega^{2}$, the Kardar-Parisi-Zhang
broadening of the 1D sound peak found in nonlinear fluctuating hydrodynamics
\cite{Spohn2014,KulkarniLamacraft2013}. For 1D electron liquids this program was carried out in Refs. 
\cite{PSMG2021,SPMG2019,GPSM2023}, whose hierarchy of
hydrodynamic regimes maps onto the two-fluid framework used here: their four-mode regime, with
the conserved quadruple of chiral densities, energy, and momentum, is the Ref. \cite{MA2019}
two-fluid window in different variables, and their three-mode regime beyond the exponentially
long equilibration scale is the single-fluid regime with anomalous coefficients. 
For the present work the important point is that the nonlinear broadening overtakes the
linear (Gaussian) theory only beyond a crossover scale parametrically longer than the onset of
the four-mode window \cite{PSMG2021}. All observables computed here are dominated by the finite
scale $q\sim1/d$ and by mode frequencies far from the deep infrared, where the Gaussian
description is controlled. Two
further circumstances support the theory. The anomalies are driven by thermal noise and weaken
rapidly with decreasing temperature. The Kardar-Parisi-Zhang mechanism relies on the resonant self-interaction of a
dispersionless sound mode, whereas the 1D plasmon is stiffened by the Coulomb logarithm, which
detunes the resonance. The fate of nonlinear fluctuating hydrodynamics with unscreened Coulomb
interactions is, to our knowledge, an open problem; the classical long-range chains studied so
far show suppressed or qualitatively altered anomalies \cite{BenentiLepriLivi2020}. The
stochastic equations constructed here are a natural starting point for its mode-coupling
analysis. Finally, the equilibration physics that controls
the dc drag is kinematic in origin and is not affected by fluctuation renormalizations: the
latter modify prefactors and not the activation law.

A remark is in order on the charge-density excitations at wave vectors near $2k_F$, created
by the vertex operators $e^{\pm2i\phi}$, which have no quasiparticle (single-boson) character.
In the bosonized description the Fock space of bosons together with the zero modes $(N,J)$ is
complete, so these operators create multi-boson states of the same excitation gas rather than an
additional branch. They do not add a slow variable, do not contribute to the thermodynamic
functions beyond the phonon equation of state used here, and carry no spectral weight in the
density response at $q\ll k_F$ and $\omega\ll vk_F$; the mode structure of
Secs.~\ref{sec:theory}--\ref{sec:dsf} is therefore unaffected. Their role is confined to the
kinetic coefficients and to the interwire coupling. Within a wire, a backscattering event, which
in bosonized language is the action of $e^{2i\phi}$ with the momentum $2k_F$ supplied by other
particles near the band bottom \cite{Lunde2007,Micklitz2010,Ristivojevic2013,Protopopov2014}, is
precisely the process that changes $J$: the $2k_F$ sector is the microscopic content of the
mutual-friction coefficient $\Gamma_J=\rho_{\rm ex}/\tau_J$ (and, in multichannel wires, of the
interchannel rate $\tau_{\rm ic}$ in Eq.~\eqref{eq:DGM}), which the hydrodynamic theory takes as
input. Between the wires, the $2k_F$ harmonic of the Coulomb coupling,
$\propto U_{12}(2k_F)\cos2(\phi_1-\phi_2)$, is the backscattering drag of
Refs.~\cite{KlesseStern2000,NazarovAverin1998}. Its amplitude is suppressed as $e^{-2k_Fd}$, but it
is a relevant perturbation for $K_-<1$, with $K_-$ the Luttinger parameter of the antisymmetric
charge sector, and grows as a power of $1/T$. In the two-fluid language it is a direct friction
link between the two zero modes, $\gamma_{\rm bs}(u_{0,1}-u_{0,2})$, which bypasses the excitation
gases and adds in parallel to the series chain of Eq.~\eqref{eq:series},
$r_D^{\rm meas}=e^{-2}n^{-2}[\gamma_{\rm bs}+\gamma_J\gamma_D/(\gamma_J+2\gamma_D)]$. The present
theory applies for $k_Fd\gg1$ and at temperatures above the scale at which $\gamma_{\rm bs}$
becomes comparable to the activated $\gamma_J$; below it the zero modes lock
\cite{KlesseStern2000} and the drag ceases to be hydrodynamic. The low-temperature upturns
reported in Refs.~\cite{Laroche2014,Makaju2025} are natural candidates for this crossover, and the
competition between an activated $\gamma_J$ and a power-law $\gamma_{\rm bs}$ is a concrete
prediction of the combined picture.

A quick remark on the roles of hydrodynamics and kinetics in this work.
Everything except the value of the backscattering time $\tau_J$ (the mode
structure, the noise algebra, the drag crossover \eqref{eq:series}, the activated law
\eqref{eq:activated} with its thermodynamic prefactor, the heat conductance
\eqref{eq:kappa12}) follows from conservation laws, thermodynamics, and the
fluctuation-dissipation theorem, with no assumption of weak intrawire interaction. What kinetic
theory alone provides is the microscopic value of $\tau_J$ (and of the fast coefficients such as
the multichannel viscosity \eqref{eq:DGM}). This nonperturbative character is the main advantage of the fluctuational approach to
one-dimensional transport, as it was for strongly correlated double layers in two dimensions
\cite{ALA2014,LLA2022,ZL2023}.

On the experimental side, the most direct arena is provided by vertically integrated 1D--1D
quantum wires, where Coulomb drag has been measured with fine control over density and
subband occupancy and where both positive and negative drag as well as low-temperature upturns
have been reported \cite{Laroche2011,Laroche2014,Yamamoto2006,Debray2001,Makaju2025,Zheng2025,Zheng2026}. Within
the present framework such devices, being short on the scale of the equilibration length, sit in
the bottlenecked regime of Eq.~\eqref{eq:series}, and the predicted observables are the activated
scale \eqref{eq:activated}, the frequency knee of Eq.~\eqref{eq:acdrag}, the drive-free
cross-wire noise spectra of Fig.~\ref{fig:S2}, and the near-field heat conductance
\eqref{eq:kappa12}, the latter two being insensitive to the bottleneck. Beyond quantum wires, the
theory applies with minor modifications to cold-atom realizations of one-dimensional liquids,
where Bragg spectroscopy accesses the structure factors directly. Among theoretical extensions we
mention the spinful case, where spin-charge separation enriches the mode content; multichannel
wires, where the coefficient set of Eq.~\eqref{eq:noise1d} is power law in temperature and the
entire theory operates without exponential restrictions; the quantitative theory of the crossover
between the collision-dominated spectra obtained here and the edge-singularity response of the
nonlinear Luttinger liquid \cite{PKKG2006,ISG2012}; and the mode-coupling analysis of the
stochastic two-fluid equations as the controlled route to anomalous hydrodynamics of charged 1D
liquids.

\begin{acknowledgments}
I am grateful to Dominique Laroche for discussions and prior collaborations on 1D Coulomb drag problems 
that in part motivated this work, and to Anton Andreev for valuable discussions regarding Ref.~\cite{MA2019}.
I also thank Dmitry Zverevich for his interest in the project and for his participation in its early stages. 
This work was financially supported by National Science Foundation (NSF) Grant No. DMR-2452658 and H. I. Romnes Faculty Fellowship provided by the University of Wisconsin-Madison Office of the Vice Chancellor for Research and Graduate Education with funding from the Wisconsin Alumni Research Foundation. This work in part employed the large language model Claude (Anthropic) for manuscript preparation. 
The project was conceptualized and technically implemented by the author, who conceived the project and carried out the research.  
\end{acknowledgments}

\appendix

\section{Fluctuational drag in the single-fluid regime: viscous channel}
\label{app:singlefluid}

In this Appendix we present the calculation of the hydrodynamic drag resistivity
$r_D^{\rm hydro}$ entering Eq.~\eqref{eq:series}, in the minimal single-fluid model with viscous
dissipation; the intrinsic-current and thermal channels are treated in
Appendixes~\ref{app:sigma} and \ref{app:thermal}.

\subsection{Model and density correlator}

Each wire is described by the linearized continuity and Navier--Stokes equations; the active wire
(labeled 1) carries the drift velocity $v$, the passive wire (labeled 2) is at rest. In Fourier
components,
\begin{align}
-i(\omega-qv)\,\delta n_1+iqn\,\delta v_1&=0,\nonumber\\
\big[-i\rho(\omega-qv)+\zeta q^{2}\big]\delta v_1&=-iqnU\big[\delta n_1+E\,\delta n_2\big]+iq\,\Xi_1,
\label{eq:app1}
\end{align}
with $E\equiv e^{-|q|d}$, and the same equations with $v=0$ and $1\leftrightarrow2$ for the
passive wire; $\langle\Xi_i\Xi_j\rangle=2\zeta T\delta_{ij}\delta\delta$. The intrinsic pressure
is neglected against the Coulomb term, which is justified with logarithmic accuracy. In the
symmetrized basis $\delta n_\pm=\delta n_1\pm\delta n_2$ the equilibrium problem decouples into
the $\pm$ sectors with the propagators \eqref{eq:Gammapm}, and the drift generates a linear-in-$v$
mixing of the sectors. Writing $G_\pm=1/\Gamma_\pm$, $\lambda=2\omega+i\omega_\nu$, and
$A=nq^{2}/\rho$, the solution to first order in $v$ reads
$\delta n_\pm^{(0)}=-AG_\pm\Xi_\pm$ and
$\delta n_\pm^{(1)}=\tfrac{qv}{2}\lambda G_\pm(\delta n_+^{(0)}+\delta n_-^{(0)})$. The
$O(v)$ part of the cross-correlator, even in frequency (the odd part integrates to zero),
assembles into
\begin{equation}
\big\langle\delta n_1\delta n_2\big\rangle^{(1)}_{\rm even}
=iqv\,\frac{T\zeta n^{2}}{\rho^{2}}\,
\frac{\omega_\nu\,q^{4}\,\big(\omega_+^{2}-\omega_-^{2}\big)}{|\Gamma_+|^{2}|\Gamma_-|^{2}} ,
\label{eq:appcorr}
\end{equation}
where we used
$\im\big[\lambda\Gamma^{*}_\pm\big]=-\omega_\nu(\omega^{2}+\omega_\pm^{2})$. 
The frequency sum rules \eqref{eq:sumrule} confirm that the Langevin normalization is
consistent with the fluctuation-dissipation theorem.

\subsection{Drag resistivity and scaling function}

\begin{figure}[t]
\includegraphics[width=\columnwidth]{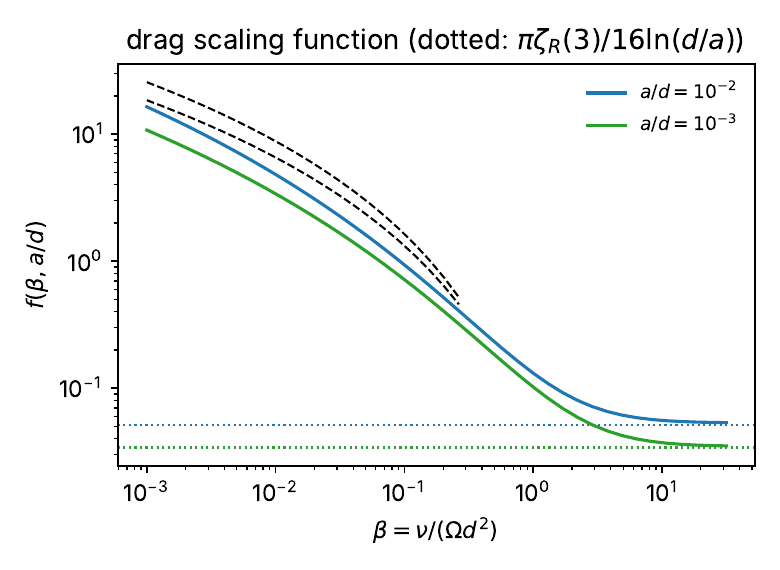}
\caption{Scaling function of the viscous drag channel, Eq.~\eqref{eq:fcorr}, for two values of
$a/d$, together with the underdamped asymptote (dashed) and the overdamped limits
$\pi\zeta_R(3)/16\ln(d/a)$ (dotted).}
\label{fig:f}
\end{figure}

The drag force per unit length exerted on the passive wire is
$\mathcal F_D=\int\!\frac{\dd\omega\,\dd q}{4\pi^{2}}(-iq)U_{12}(q)
\langle\delta n_1\delta n_2\rangle$, and $r_D=\mathcal F_D/e^{2}n^{2}v$. Substituting
Eq.~\eqref{eq:appcorr} and using $\omega_+^{2}-\omega_-^{2}=2\omega_p^{2}E$,
\begin{equation}
r_D^{(\zeta)}=\frac{4T\zeta}{\pi^{2}\rho^{2}}\int_0^\infty\!\!\dd q\!\int_0^\infty\!\!\dd\omega\,
\ln\left(\frac{1}{aq}\right)\,e^{-2qd}
\frac{\omega_\nu\,\omega_p^{2}\,q^{6}}{|\Gamma_+|^{2}|\Gamma_-|^{2}} .
\end{equation}
Rescaling $q=x/d$ and $\omega=\Omega y$ with $\Omega^{2}=2e^{2}n^{2}/\rho d^{2}$, the frequency
integral is elementary though lengthy; for arbitrary mode splitting and damping it evaluates in
closed form to
\begin{align}
&\int_0^\infty\frac{\dd y}{\big[(y^{2}-a_1^{2})^{2}+b^{2}y^{2}\big]\big[(y^{2}-a_2^{2})^{2}+b^{2}y^{2}\big]}
\nonumber\\
&\qquad
=\frac{\pi}{2b\,a_1^{2}a_2^{2}}\,
\frac{a_1^{2}+a_2^{2}+2b^{2}}{(a_1^{2}-a_2^{2})^{2}+2b^{2}(a_1^{2}+a_2^{2})} ,
\label{eq:appI}
\end{align}
with $a_{1,2}^{2}=x^{2}L(x)(1\pm e^{-x})$, $L(x)=\ln[d/(ax)]$, and $b=\beta x^{2}$,
$\beta=\nu/\Omega d^{2}$, $\nu=\zeta/\rho$. Carrying out the substitution of the factor
$a_1^{2}a_2^{2}=x^{4}L^{2}(1-e^{-2x})$ and of the logarithms in the second denominator of
Eq.~\eqref{eq:appI}, one arrives at the result quoted in Eq.~\eqref{eq:rdvisc} with the scaling
function
\begin{equation}
f\Big(\beta,\frac{a}{d}\Big)=\frac{\pi}{4}\int_0^{d/a}\!\!\dd x\,
\frac{x^{2}e^{-2x}\,\big[L(x)+\beta^{2}x^{2}\big]}
{L(x)\big[1-e^{-2x}\big]\big[L(x)e^{-2x}+\beta^{2}x^{2}\big]} .
\label{eq:fcorr}
\end{equation}
The function is plotted in Fig.~\ref{fig:f}. In the
underdamped regime, $\beta\ll1$, the interwire exponential cancels between the numerator and the
mode-splitting term of the denominator: the drag is dominated by nearly degenerate pairs of
plasmon resonances over the broad interval $1\lesssim x\lesssim x_c=\ln(\sqrt\Lambda/\beta)$, and
with logarithmic accuracy
\begin{equation}
f\simeq\frac{\pi}{12}\,\frac{\ln^{3}\!\big(\sqrt{\Lambda}/\beta\big)}{\Lambda},
\qquad \Lambda=\ln\frac{d}{a},
\label{eq:fasym}
\end{equation}
the one-dimensional counterpart of the $\ln^{4}$ plasmon enhancement of the bilayer drag
\cite{ALA2014}. In the overdamped limit, $\beta\gg1$,
$f\to\pi\zeta_R(3)/16\Lambda$. In both limits $r_D^{(\zeta)}\propto T\zeta$ up to logarithms: the
hydrodynamic drag measures the bulk viscosity of the wire.

\section{Intrinsic-current channel and the hydrodynamic origin of the activated drag}
\label{app:sigma}

In the single-fluid regime the currents \eqref{eq:MAgamma} carry, alongside the viscous stress,
the Langevin flux of the intrinsic-current channel, Eq.~\eqref{eq:Upsilonnoise}. Retaining the
$\gamma_{11}$ entry (the thermoelectric channel couples to the thermal sector treated in
Appendix~\ref{app:thermal}), the continuity equation of each wire acquires the dissipative and
stochastic terms
\begin{equation}
\partial_t\delta n+\partial_x\big[n\delta v
-\gamma_{11}\partial_x\big(\delta\mu+e\delta\Phi\big)+\xi_n\big]=0,
\end{equation}
the one-dimensional transplant of the construction of Ref.~\cite{ZL2023}. With logarithmic
accuracy the electrochemical gradient is dominated by the Coulomb term, and the charge
fluctuations acquire the Maxwell relaxation rates
$\gamma_\pm=\gamma_{11}q^{2}U(q)(1\pm E)$ alongside the plasmon frequencies; the propagators
generalize to
$\tilde\Gamma_\pm=\omega^{2}-\omega_\pm^{2}-\omega_\nu\gamma_\pm+i\omega(\omega_\nu+\gamma_\pm)$.
Repeating the calculation of Appendix~\ref{app:singlefluid} with the two uncorrelated noise
families, the $O(v)$ frequency-even cross-correlator becomes
\begin{align}
\big\langle\delta n_1\delta n_2\big\rangle^{(1)}_{\rm even}
=\frac{iqvT}{|\tilde\Gamma_+|^{2}|\tilde\Gamma_-|^{2}}
\left\{\frac{n^{2}q^{4}\zeta}{\rho^{2}}
\Big[\omega_\nu\Delta_p+\big(3\omega^{2}+\omega_\nu^{2}\big)\Delta_\gamma\Big]\right. \nonumber \\ 
\left.+q^{2}\gamma_{11}\Big[\big(\omega^{2}+\omega_\nu^{2}\big)^{2}\Delta_\gamma
+\omega_\nu\big(\omega_\nu^{2}-\omega^{2}\big)\Delta_p\Big]\right\},
\label{eq:twochannel}
\end{align}
with $\Delta_p=\omega_+^{2}-\omega_-^{2}=2\omega_p^{2}E$ and
$\Delta_\gamma=\gamma_+-\gamma_-=2\gamma_{11}q^{2}UE$. Equation \eqref{eq:twochannel} reproduces the structure
$\propto\omega^{4}\Delta_\gamma$ of the bilayer intrinsic-current mechanism \cite{ZL2023}. Note
the identity $\gamma_+\omega_-^{2}=\gamma_-\omega_+^{2}$, both splittings inherit the same
screening factor $1\pm E$, which eliminates the naively leading cross terms, exactly as in two
dimensions.

\begin{figure}[t]
\includegraphics[width=\columnwidth]{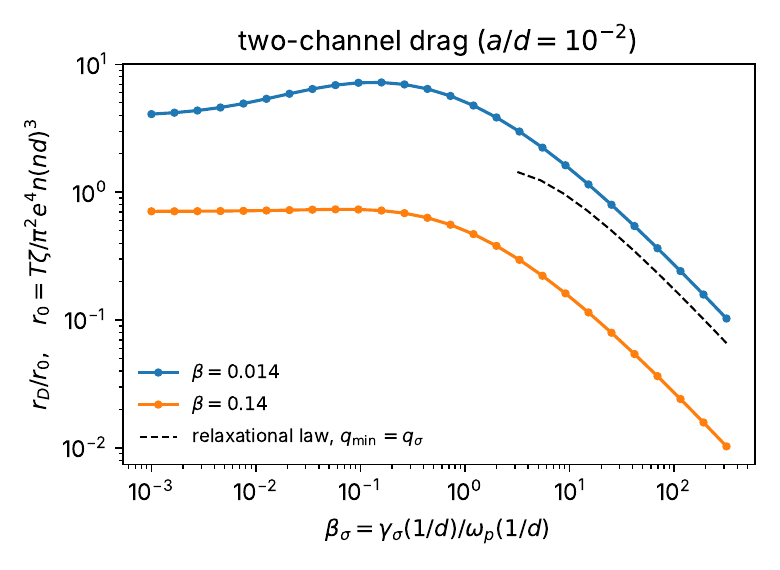}
\caption{Two-channel drag versus the Maxwell damping ratio
$\beta_\sigma=\gamma_\sigma(1/d)/\omega_p(1/d)$, in units of the viscous drag scale
$r_0=T\zeta/\pi^{2}e^{4}n(nd)^{3}$, for two values of the viscous damping ratio $\beta$ and
$a/d=10^{-2}$. The plateau at small $\beta_\sigma$ equals the scaling function $f(\beta,a/d)$ of
Eq.~\eqref{eq:fcorr}; the relaxational regime follows Eq.~\eqref{eq:rdsigma} with
$q_{\min}=q_\sigma$ (dashed), $r_D\propto1/\beta_\sigma$.}
\label{fig:sigma}
\end{figure}

The resulting drag is shown in Fig.~\ref{fig:sigma} as a function of the Maxwell damping ratio
$\beta_\sigma=\gamma_\sigma(1/d)/\omega_p(1/d)$, with
$\gamma_\sigma=\gamma_{11}q^{2}U$. At small $\beta_\sigma$ the viscous plateau of
Appendix~\ref{app:singlefluid} is recovered; once $\gamma_\pm(1/d)\gtrsim\omega_\pm(1/d)$ the
charge fluctuations obey Maxwell relaxation rather than plasmon oscillation, and
\begin{equation}
r_D^{(\sigma)}\simeq\frac{T}{4\pi e^{2}n^{2}\gamma_{11}}
\int_{q_{\rm min}}\!\!\dd q\,\frac{e^{-2qd}}{U(q)\big(1-e^{-2qd}\big)}
\;\propto\;\frac{1}{\gamma_{11}},
\label{eq:rdsigma}
\end{equation}
where the infrared logarithm is cut by the wire length or by the scale
$q_\sigma\simeq n/\gamma_{11}\sqrt{\rho U}$, below which $\gamma_\pm(q)\propto q^{2}$ falls under
$\omega_\pm(q)\propto q$ and Maxwell relaxation gives way to plasmon oscillation, whichever is
larger; for an infinite wire $q_{\min}=q_\sigma$ and $r_D^{(\sigma)}\propto\gamma_{11}^{-1}$ up
to a logarithm of $\gamma_{11}$. This is the one-dimensional analog of the charge-neutrality bilayer result
$k_{\rm th}\propto T/\sigma d^{3}$ of Ref.~\cite{LLA2022}: drag inversely proportional to the
intrinsic conductivity.

Two consistency conditions govern the use of Eq.~\eqref{eq:rdsigma}. The static formula samples
frequencies up to $\gamma_\pm$ and therefore requires $\gamma_\pm(1/d)\tau_J\ll1$, a condition
attainable in multichannel wires with fast zero-mode relaxation, where Eq.~\eqref{eq:rdsigma} is
the controlled dc result in its overdamped window. In a strictly single-channel wire, where
$\gamma_{11}=\tau_J\rho_{\rm ex}(1/m^{*}-1/m_0)^{2}$ is itself exponentially large, the
dispersive form \eqref{eq:debye} must be used instead. Doing so, the low-frequency region $|\omega|\lesssim\tau_J^{-1}$ contributes only at
order $1/\tau_J^{3}$, while the dominant contribution comes from the plasmon resonances, whose
backscattering-induced width and noise weight are both governed by
$\re\gamma_{11}(\omega_\pm)\simeq\gamma_{11}/(\omega_\pm\tau_J)^{2}$. Near a plasmon pole the
Maxwell term acts exactly as a viscous damping: comparing $\tilde\Gamma_\pm$ with $\Gamma_\pm$,
and the noise weights $q^{2}(\omega^{2}+\omega_\nu^{2})\,2T\re\gamma_{11}(\omega)$ and
$(n^{2}q^{4}/\rho^{2})\,2T\zeta$ at $\omega=\omega_\pm$, both map onto the viscous channel with
the effective, backscattering-limited viscosity
$\zeta_{\rm eff}(q)=\rho U(q)(1\pm E)\re\gamma_{11}(\omega_\pm)=(m_0/m^{*}-1)^{2}\gamma_J/q^{2}$.
The resonant part of the intrinsic channel is therefore Eq.~\eqref{eq:rdvisc} with
$\zeta\to\zeta_{\rm eff}$, with the outcome
\begin{equation}
r_D^{(\sigma)}\;\sim\;\Big(\frac{m_0}{m^{*}}-1\Big)^{2}\frac{\gamma_J}{e^{2}n^{2}}\;
\frac{T\tilde f}{\pi^{2}e^{2}n^{2}d}\;\propto\;e^{-D/T},
\label{eq:resonantsigma}
\end{equation}
where $\tilde f$ is the scaling function \eqref{eq:fcorr} evaluated with
$\beta\to\beta_{\rm eff}(x)=\beta_0/x^{2}$, $\beta_0=\zeta_{\rm eff}(1/d)/\rho\Omega d^{2}\propto\gamma_J$.
The intrinsic channel thus reproduces the activated law of the bottlenecked limit
\eqref{eq:activated}, with the non-Galilean weight and a non-exponential prefactor that contains
the Coulomb logarithms. The two results are not in competition: the dispersive single-fluid
calculation yields the plasmon-mediated transfer rate $\gamma_D^{(\sigma)}=e^{2}n^{2}r_D^{(\sigma)}$,
which in a single-channel wire is itself activated, and the measured drag follows from the series
law \eqref{eq:series} with both $\gamma_J$ and $\gamma_D$ proportional to $e^{-D/T}$. The apparent
paradox that exponentially large static coefficients would freeze the fluctuations and lock the
wires does not arises: the coefficients disperse before the locking regime is reached. In other
words, slow internal relaxation implies large but strongly dispersive kinetic coefficients, and
this dispersion is what protects the consistency of the theory.

These results fix the bookkeeping of the fluctuational channels with respect to the bottleneck
formula \eqref{eq:series}. Only the fluctuation transfer carried at frequencies
$\omega\tau_J\gg1$, by plasmons and second sound, passes through the zero-mode bottleneck and
belongs inside $\gamma_D$; transfer carried at $\omega\tau_J\lesssim1$ sees a locally
equilibrated liquid, bypasses the bottleneck, and adds in parallel. The suppressions
$1/\gamma_{11}\propto e^{-D/T}$ [Eq.~\eqref{eq:rdsigma}] and $1/\kappa\propto e^{-D/T}$
[Appendix~\ref{app:thermal}] appearing in the low-frequency channels are the bottleneck,
expressed in single-fluid language; multiplying them by the series factor of
Eq.~\eqref{eq:series} would count the same exponential twice. In the multichannel window
(W1)--(W3) the series factor is close to unity and the measured dc drag is the sum of the three
channels, $r_D^{\rm meas}\simeq r_D^{(\zeta)}+r_D^{(\rm th)}+r_D^{(\sigma)}$, each evaluated with
static coefficients; there the intrinsic channel is not automatically subleading---its weight
relative to the viscous one is $\gamma_\sigma(1/d)/\omega_\nu(1/d)=\gamma_{11}U\rho/\zeta$, which
can be large since $\gamma_{11}$ inherits the slowest surviving relaxation time, mirroring the
bilayer case where the intrinsic mechanism decays as $1/d^{4}$ against $1/d^{5}$ for the viscous
one \cite{ZL2023}. In a single-channel wire all three channels are activated and comparable,
with the distinct thermodynamic prefactors derived here and in Appendix~\ref{app:thermal}.

\section{Thermal-flux channel}
\label{app:thermal}

\begin{figure}[t]
\includegraphics[width=\columnwidth]{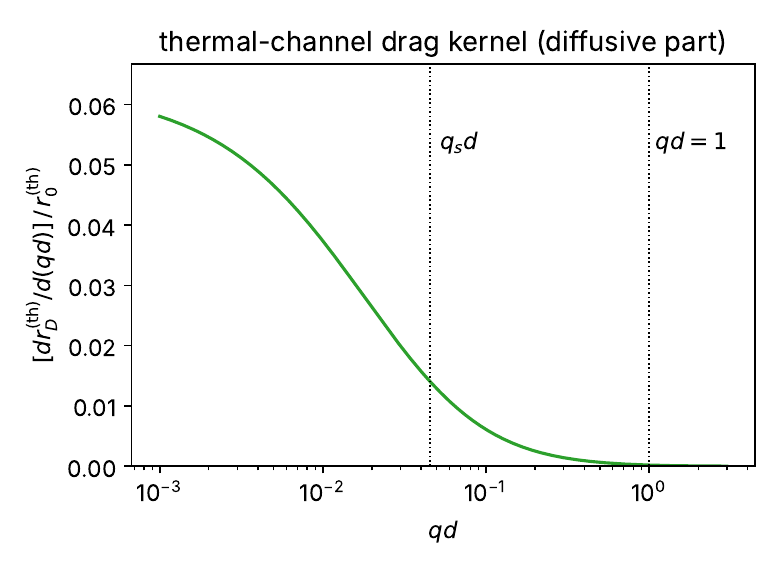}
\caption{Thermal-channel drag kernel, the integrand of Eq.~\eqref{eq:rdthermal} per unit
$qd$, normalized by the scale $r_0^{(\rm th)}=T^{3}(\partial s/\partial\ln n)_T^{4}/
e^{4}n^{2}mc\,\kappa\,d\,v_s^{2}$ of Eq.~\eqref{eq:rdthermalestimate}. The kernel is dominated by the
sound--plasmon crossover scale $q_s$, far below the interwire scale $1/d$ that dominates the
viscous and intrinsic channels.}
\label{fig:th}
\end{figure}

For completeness we establish the one-dimensional form of the thermal-flux contribution to the
hydrodynamic drag, the analog of the thermal mechanism that dominates the bilayer result
\cite{ALA2014}. Restoring the entropy sector, each wire is described by the fields
$(\delta n,\delta v,\delta s)$, with $s$ the entropy per particle:
\begin{align}
&\partial_t\delta n+n\partial_x\delta v=0,\nonumber\\
&\rho\,\partial_t\delta v=-\partial_x\Big[\Big(\tfrac{\partial P}{\partial n}\Big)_{\!s}\delta n
+\Big(\tfrac{\partial P}{\partial s}\Big)_{\!n}\delta s+ne\delta\Phi\Big]
+\zeta\partial_x^{2}\delta v+\partial_x\Xi,\nonumber\\
&nT\,\partial_t\delta s=\kappa\,\partial_x^{2}\delta T-\partial_xg,\qquad
\langle gg\rangle=2\kappa T^{2}\delta\delta,
\end{align}
with $\delta T=(\partial T/\partial n)_s\delta n+(\partial T/\partial s)_n\delta s$ and the
thermodynamic identity $(\partial P/\partial s)_n=n^{2}(\partial T/\partial n)_s$. At frequencies
of the order of the thermal diffusion rate the momentum balance is quasistatic: the
Coulomb-stiffened plasmon screens the entropy-induced charge,
$\delta n_\pm=-\Lambda_\pm\delta s_\pm$ with
$\Lambda_\pm=nq^{2}(\partial P/\partial s)_n/\rho\omega_\pm^{2}$, and the heat then diffuses with
the screened rates
\begin{equation}
\tilde\gamma_\pm=D_Tq^{2}\Big(1-\frac{\omega_\alpha^{2}}{\omega_\pm^{2}}\Big),\quad
\omega_\alpha=u_{\rm th}q,\quad
u_{\rm th}^{2}=\frac{T}{mc}\Big(\frac{\partial s}{\partial\ln n}\Big)_{\!T}^{2},
\end{equation}
where $D_T=\kappa/nc$ is the thermal diffusivity and $c$ is the specific heat per particle: this is the familiar shift
between constant-density and constant-pressure heat diffusion, operating here under the screening
constraint. It is essential that
$\omega_\pm^{2}=[n^{2}U(1\pm E)+nmv_s^{2}]q^{2}/\rho$ retains the intrinsic-pressure contribution
$mv_s^{2}=(\partial P/\partial n)_s$, with $v_s$ the sound velocity, which controls the infrared behavior. The frequency-even
$O(v)$ cross-correlator of this sector is
\begin{equation}
\big\langle\delta n_1\delta n_2\big\rangle^{(1),\rm th}_{\rm even}
=iqv\,\frac{\Lambda_+\Lambda_-\,\kappa q^{2}\,(\tilde\gamma_+-\tilde\gamma_-)}
{n^{2}\,(\tilde\gamma_+^{2}+\omega^{2})(\tilde\gamma_-^{2}+\omega^{2})}.
\label{eq:thermalcorr}
\end{equation}
The frequency integral is elementary and gives
\begin{equation}
r_D^{(\rm th)}=\frac{1}{2\pi e^{2}n^{2}}\int_0^\infty\!\!\dd q\,
q^{2}U(q)e^{-qd}
\frac{\Lambda_+\Lambda_-\kappa q^{2}(\tilde\gamma_+-\tilde\gamma_-)}
{n^{2}\tilde\gamma_+\tilde\gamma_-(\tilde\gamma_++\tilde\gamma_-)} .
\label{eq:rdthermal}
\end{equation}
Three properties follow from Eq.~\eqref{eq:rdthermal}: it is proportional to
$1/\kappa$; it scales as the fourth power of the thermal-expansion derivative,
$r_D^{(\rm th)}\propto(\partial s/\partial\ln n)_T^{4}$, up to a logarithmic drift of the
dominant scale; and the integral is dominated by the sound--plasmon crossover scale 
$q_s\simeq mv_s^{2}/[nU(q_s)d]\ll1/d$, where the antisymmetric acoustic plasmon
softens into ordinary sound and the screening of the entropy-induced charge is weakest
(Fig.~\ref{fig:th}). Evaluating the integral at the dominant scale $q_s$, where the
integrand is flat for $q<q_s$ and falls off as $q^{-2}$ above, gives, up to a logarithmic factor
$\sim[16\pi\ln(1/aq_s)]^{-1}$,
\begin{equation}
r_D^{(\rm th)}\sim\frac{T^{3}}{e^{4}n^{2}mc\,\kappa\,d\,v_s^{2}}
\Big(\frac{\partial s}{\partial\ln n}\Big)_{\!T}^{4},
\label{eq:rdthermalestimate}
\end{equation}
the one-dimensional descendant of the thermal term of the bilayer drag \cite{ALA2014}: inversely
proportional to the thermal conductivity and quartic in the thermal expansion, but falling off
only as $1/d$, because the transfer is dominated by wavelengths of order $1/q_s\gg d$. In a
single-channel wire $\kappa=\pi\tau_JTv/3\hbar$ \cite{MR2015}, so
$r_D^{(\rm th)}\propto e^{-D/T}$: the third independent hydrodynamic route to the activation law,
completing the uniform picture of the channel bookkeeping in Appendix~\ref{app:sigma}.

\section{Momentum balance for coupled wires}
\label{app:balance}

Let $u_{0,i}$ and $u_i$ denote the zero-mode and excitation-gas drift velocities of wire $i$, let
$\gamma_J$ be the zero-mode--excitation friction coefficient \eqref{eq:GammaJ}, and let
$\gamma_D$ be the interwire friction coefficient defined through the fluctuational drag force
$F_D=\gamma_D(u_1-u_2)$; comparison with Appendix~\ref{app:singlefluid} identifies
$\gamma_D=e^{2}n^{2}r_D^{\rm hydro}$ in the fully equilibrated limit. A uniform electric field
acts on the total charge, and since long-wavelength excitations are neutral density waves, the
field force enters the zero-mode equations only. With the drive $u_{0,1}=v$ maintained by the
current source (at $T\ll T_F$ the current is carried almost entirely by the zero mode) and the
passive wire open-circuited ($u_{0,2}\approx0$ to leading order in $\rho_{\rm ex}/\rho$), so
that the electric field $E_2$ induced in the passive wire balances the mutual friction exerted
on its stationary zero mode by the drifting excitation gas, $enE_2=\gamma_Ju_2$, the
stationary balances read
\begin{align}
\gamma_J(v-u_1)&=\gamma_D(u_1-u_2),\nonumber\\
\gamma_D(u_1-u_2)&=\gamma_J u_2,\qquad
enE_2=\gamma_J u_2\;[=\gamma_D(u_1-u_2)],
\end{align}
with the solution
\begin{equation}
u_1=v\,\frac{\gamma_J+\gamma_D}{\gamma_J+2\gamma_D},\qquad
E_2=\frac{\gamma_D}{en}\,\frac{\gamma_J\,v}{\gamma_J+2\gamma_D},
\end{equation}
which yields Eq.~\eqref{eq:series} with $r_D^{\rm meas}=E_2/env$. The factor $\gamma_J/(\gamma_J+2\gamma_D)$ is the fraction of
the drive velocity actually acquired by the excitation gas of the active wire, i.e., the degree
to which the thermal fluctuations are entrained by the current; the factor of 2 counts the two
$\gamma_J$ links of the series chain. In the limit $\gamma_J\ll\gamma_D$ of infinite wires the
two excitation gases, strongly coupled by $\gamma_D$, drift together at
$u_1\simeq u_2\simeq v/2$, and the voltage is nevertheless $\gamma_J$-limited because the last
link of the chain is; in wires shorter than $\ell_J$ the excitation gases are anchored to the
stationary leads instead, and the entrainment itself never develops. Retaining the inertia of the
excitation gases, $\rho_{\rm ex}\dot u_i$, in the same equations yields the finite-frequency
result \eqref{eq:acdrag}.

\bibliography{biblio-TwoFluidLL}

\begin{thebibliography}{64}%
\makeatletter
\providecommand \@ifxundefined [1]{%
 \@ifx{#1\undefined}
}%
\providecommand \@ifnum [1]{%
 \ifnum #1\expandafter \@firstoftwo
 \else \expandafter \@secondoftwo
 \fi
}%
\providecommand \@ifx [1]{%
 \ifx #1\expandafter \@firstoftwo
 \else \expandafter \@secondoftwo
 \fi
}%
\providecommand \natexlab [1]{#1}%
\providecommand \enquote  [1]{``#1''}%
\providecommand \bibnamefont  [1]{#1}%
\providecommand \bibfnamefont [1]{#1}%
\providecommand \citenamefont [1]{#1}%
\providecommand \href@noop [0]{\@secondoftwo}%
\providecommand \href [0]{\begingroup \@sanitize@url \@href}%
\providecommand \@href[1]{\@@startlink{#1}\@@href}%
\providecommand \@@href[1]{\endgroup#1\@@endlink}%
\providecommand \@sanitize@url [0]{\catcode `\\12\catcode `\$12\catcode
  `\&12\catcode `\#12\catcode `\^12\catcode `\_12\catcode `\%12\relax}%
\providecommand \@@startlink[1]{}%
\providecommand \@@endlink[0]{}%
\providecommand \url  [0]{\begingroup\@sanitize@url \@url }%
\providecommand \@url [1]{\endgroup\@href {#1}{\urlprefix }}%
\providecommand \urlprefix  [0]{URL }%
\providecommand \Eprint [0]{\href }%
\providecommand \doibase [0]{http://dx.doi.org/}%
\providecommand \selectlanguage [0]{\@gobble}%
\providecommand \bibinfo  [0]{\@secondoftwo}%
\providecommand \bibfield  [0]{\@secondoftwo}%
\providecommand \translation [1]{[#1]}%
\providecommand \BibitemOpen [0]{}%
\providecommand \bibitemStop [0]{}%
\providecommand \bibitemNoStop [0]{.\EOS\space}%
\providecommand \EOS [0]{\spacefactor3000\relax}%
\providecommand \BibitemShut  [1]{\csname bibitem#1\endcsname}%
\let\auto@bib@innerbib\@empty
\bibitem [{\citenamefont {Landau}\ and\ \citenamefont
  {Lifshitz}(1957)}]{LandauLifshitz1957}%
  \BibitemOpen
  \bibfield  {author} {\bibinfo {author} {\bibfnamefont {L.~D.}\ \bibnamefont
  {Landau}}\ and\ \bibinfo {author} {\bibfnamefont {E.~M.}\ \bibnamefont
  {Lifshitz}},\ }\bibfield  {title} {\enquote {\bibinfo {title} {Hydrodynamic
  fluctuations},}\ }\href@noop {} {\bibfield  {journal} {\bibinfo  {journal}
  {Zh. Eksp. Teor. Fiz.}\ }\textbf {\bibinfo {volume} {32}},\ \bibinfo {pages}
  {618} (\bibinfo {year} {1957})},\ \bibinfo {note} {[Sov. Phys. JETP {\bf 5},
  512 (1957)]}\BibitemShut {NoStop}%
\bibitem [{\citenamefont {Lifshitz}\ and\ \citenamefont
  {Pitaevskii}(1980)}]{LL9}%
  \BibitemOpen
  \bibfield  {author} {\bibinfo {author} {\bibfnamefont {E.~M.}\ \bibnamefont
  {Lifshitz}}\ and\ \bibinfo {author} {\bibfnamefont {L.~P.}\ \bibnamefont
  {Pitaevskii}},\ }\href@noop {} {\emph {\bibinfo {title} {Statistical Physics,
  Part 2}}},\ \bibinfo {series} {Course of Theoretical Physics}, Vol.~\bibinfo
  {volume} {9}\ (\bibinfo  {publisher} {Pergamon Press},\ \bibinfo {address}
  {New York},\ \bibinfo {year} {1980})\BibitemShut {NoStop}%
\bibitem [{\citenamefont {Callen}\ and\ \citenamefont
  {Welton}(1951)}]{CallenWelton1951}%
  \BibitemOpen
  \bibfield  {author} {\bibinfo {author} {\bibfnamefont {Herbert~B.}\
  \bibnamefont {Callen}}\ and\ \bibinfo {author} {\bibfnamefont {Theodore~A.}\
  \bibnamefont {Welton}},\ }\bibfield  {title} {\enquote {\bibinfo {title}
  {Irreversibility and generalized noise},}\ }\href {\doibase
  10.1103/PhysRev.83.34} {\bibfield  {journal} {\bibinfo  {journal} {Physical
  Review}\ }\textbf {\bibinfo {volume} {83}},\ \bibinfo {pages} {34--40}
  (\bibinfo {year} {1951})}\BibitemShut {NoStop}%
\bibitem [{\citenamefont {Khalatnikov}(1957)}]{Khalatnikov1957}%
  \BibitemOpen
  \bibfield  {author} {\bibinfo {author} {\bibfnamefont {I.~M.}\ \bibnamefont
  {Khalatnikov}},\ }\bibfield  {title} {\enquote {\bibinfo {title}
  {Hydrodynamic fluctuations in a superfluid liquid},}\ }\href@noop {}
  {\bibfield  {journal} {\bibinfo  {journal} {Zh. Eksp. Teor. Fiz.}\ }\textbf
  {\bibinfo {volume} {33}},\ \bibinfo {pages} {809} (\bibinfo {year} {1957})},\
  \bibinfo {note} {[Sov. Phys. JETP {\bf 6}, 624 (1958)]}\BibitemShut {NoStop}%
\bibitem [{\citenamefont {Khalatnikov}(1965)}]{KhalatnikovBook}%
  \BibitemOpen
  \bibfield  {author} {\bibinfo {author} {\bibfnamefont {I.~M.}\ \bibnamefont
  {Khalatnikov}},\ }\href@noop {} {\emph {\bibinfo {title} {An Introduction to
  the Theory of Superfluidity}}}\ (\bibinfo  {publisher} {Benjamin},\ \bibinfo
  {address} {New York},\ \bibinfo {year} {1965})\BibitemShut {NoStop}%
\bibitem [{\citenamefont {Onsager}(1931)}]{Onsager1931}%
  \BibitemOpen
  \bibfield  {author} {\bibinfo {author} {\bibfnamefont {Lars}\ \bibnamefont
  {Onsager}},\ }\bibfield  {title} {\enquote {\bibinfo {title} {Reciprocal
  relations in irreversible processes. {I}.}}\ }\href {\doibase
  10.1103/PhysRev.37.405} {\bibfield  {journal} {\bibinfo  {journal} {Phys.
  Rev.}\ }\textbf {\bibinfo {volume} {37}},\ \bibinfo {pages} {405--426}
  (\bibinfo {year} {1931})}\BibitemShut {NoStop}%
\bibitem [{\citenamefont {Andreev}(1970)}]{Andreev1971}%
  \BibitemOpen
  \bibfield  {author} {\bibinfo {author} {\bibfnamefont {A.~F.}\ \bibnamefont
  {Andreev}},\ }\bibfield  {title} {\enquote {\bibinfo {title} {Two-liquid
  effects in a normal liquid},}\ }\href@noop {} {\bibfield  {journal} {\bibinfo
   {journal} {Zh. Eksp. Teor. Fiz.}\ }\textbf {\bibinfo {volume} {59}},\
  \bibinfo {pages} {1819} (\bibinfo {year} {1970})},\ \bibinfo {note} {[Sov.
  Phys. JETP {\bf 32}, 987 (1971)]}\BibitemShut {NoStop}%
\bibitem [{\citenamefont {Gurzhi}(1968)}]{Gurzhi1968}%
  \BibitemOpen
  \bibfield  {author} {\bibinfo {author} {\bibfnamefont {R.~N.}\ \bibnamefont
  {Gurzhi}},\ }\bibfield  {title} {\enquote {\bibinfo {title} {Hydrodynamic
  effects in solids at low temperature},}\ }\href@noop {} {\bibfield  {journal}
  {\bibinfo  {journal} {Usp. Fiz. Nauk}\ }\textbf {\bibinfo {volume} {94}},\
  \bibinfo {pages} {689} (\bibinfo {year} {1968})},\ \bibinfo {note} {[Sov.
  Phys. Usp. {\bf 11}, 255 (1968)]}\BibitemShut {NoStop}%
\bibitem [{\citenamefont {Gurzhi}\ and\ \citenamefont
  {Kopeliovich}(1981)}]{GurzhiKopeliovich}%
  \BibitemOpen
  \bibfield  {author} {\bibinfo {author} {\bibfnamefont {R.~N.}\ \bibnamefont
  {Gurzhi}}\ and\ \bibinfo {author} {\bibfnamefont {A.~I.}\ \bibnamefont
  {Kopeliovich}},\ }\bibfield  {title} {\enquote {\bibinfo {title}
  {Low-temperature electrical conductivity of pure metals},}\ }\href@noop {}
  {\bibfield  {journal} {\bibinfo  {journal} {Usp. Fiz. Nauk}\ }\textbf
  {\bibinfo {volume} {133}},\ \bibinfo {pages} {33} (\bibinfo {year} {1981})},\
  \bibinfo {note} {[Sov. Phys. Usp. {\bf 24}, 17 (1981)]}\BibitemShut {NoStop}%
\bibitem [{\citenamefont {Andreev}\ \emph {et~al.}(2011)\citenamefont
  {Andreev}, \citenamefont {Kivelson},\ and\ \citenamefont {Spivak}}]{AKS2011}%
  \BibitemOpen
  \bibfield  {author} {\bibinfo {author} {\bibfnamefont {A.~V.}\ \bibnamefont
  {Andreev}}, \bibinfo {author} {\bibfnamefont {S.~A.}\ \bibnamefont
  {Kivelson}}, \ and\ \bibinfo {author} {\bibfnamefont {B.}~\bibnamefont
  {Spivak}},\ }\bibfield  {title} {\enquote {\bibinfo {title} {Hydrodynamic
  description of transport in strongly correlated electron systems},}\
  }\href@noop {} {\bibfield  {journal} {\bibinfo  {journal} {Phys. Rev. Lett.}\
  }\textbf {\bibinfo {volume} {106}},\ \bibinfo {pages} {256804} (\bibinfo
  {year} {2011})}\BibitemShut {NoStop}%
\bibitem [{\citenamefont {Haldane}(1981)}]{Haldane1981}%
  \BibitemOpen
  \bibfield  {author} {\bibinfo {author} {\bibfnamefont {F.~D.~M.}\
  \bibnamefont {Haldane}},\ }\bibfield  {title} {\enquote {\bibinfo {title}
  {'{L}uttinger liquid theory' of one-dimensional quantum fluids. {I}.
  {P}roperties of the {L}uttinger model and their extension to the general 1{D}
  interacting spinless {F}ermi gas},}\ }\href@noop {} {\bibfield  {journal}
  {\bibinfo  {journal} {J. Phys. C: Solid State Phys.}\ }\textbf {\bibinfo
  {volume} {14}},\ \bibinfo {pages} {2585} (\bibinfo {year}
  {1981})}\BibitemShut {NoStop}%
\bibitem [{\citenamefont {Imambekov}\ \emph {et~al.}(2012)\citenamefont
  {Imambekov}, \citenamefont {Schmidt},\ and\ \citenamefont
  {Glazman}}]{ISG2012}%
  \BibitemOpen
  \bibfield  {author} {\bibinfo {author} {\bibfnamefont {A.}~\bibnamefont
  {Imambekov}}, \bibinfo {author} {\bibfnamefont {T.~L.}\ \bibnamefont
  {Schmidt}}, \ and\ \bibinfo {author} {\bibfnamefont {L.~I.}\ \bibnamefont
  {Glazman}},\ }\bibfield  {title} {\enquote {\bibinfo {title} {One-dimensional
  quantum liquids: {B}eyond the {L}uttinger liquid paradigm},}\ }\href@noop {}
  {\bibfield  {journal} {\bibinfo  {journal} {Rev. Mod. Phys.}\ }\textbf
  {\bibinfo {volume} {84}},\ \bibinfo {pages} {1253} (\bibinfo {year}
  {2012})}\BibitemShut {NoStop}%
\bibitem [{\citenamefont {Levchenko}\ and\ \citenamefont
  {Micklitz}(2021)}]{LevchenkoMicklitz2021}%
  \BibitemOpen
  \bibfield  {author} {\bibinfo {author} {\bibfnamefont {A.}~\bibnamefont
  {Levchenko}}\ and\ \bibinfo {author} {\bibfnamefont {T.}~\bibnamefont
  {Micklitz}},\ }\bibfield  {title} {\enquote {\bibinfo {title} {Kinetic
  processes in {F}ermi--{L}uttinger liquids},}\ }\href@noop {} {\bibfield
  {journal} {\bibinfo  {journal} {J. Exp. Theor. Phys.}\ }\textbf {\bibinfo
  {volume} {132}},\ \bibinfo {pages} {675} (\bibinfo {year}
  {2021})}\BibitemShut {NoStop}%
\bibitem [{\citenamefont {Lunde}\ \emph {et~al.}(2007)\citenamefont {Lunde},
  \citenamefont {Flensberg},\ and\ \citenamefont {Glazman}}]{Lunde2007}%
  \BibitemOpen
  \bibfield  {author} {\bibinfo {author} {\bibfnamefont {Anders~Mathias}\
  \bibnamefont {Lunde}}, \bibinfo {author} {\bibfnamefont {Karsten}\
  \bibnamefont {Flensberg}}, \ and\ \bibinfo {author} {\bibfnamefont
  {Leonid~I.}\ \bibnamefont {Glazman}},\ }\bibfield  {title} {\enquote
  {\bibinfo {title} {Three-particle collisions in quantum wires: Corrections to
  thermopower and conductance},}\ }\href {\doibase 10.1103/PhysRevB.75.245418}
  {\bibfield  {journal} {\bibinfo  {journal} {Phys. Rev. B}\ }\textbf {\bibinfo
  {volume} {75}},\ \bibinfo {pages} {245418} (\bibinfo {year}
  {2007})}\BibitemShut {NoStop}%
\bibitem [{\citenamefont {Micklitz}\ \emph {et~al.}(2010)\citenamefont
  {Micklitz}, \citenamefont {Rech},\ and\ \citenamefont
  {Matveev}}]{Micklitz2010}%
  \BibitemOpen
  \bibfield  {author} {\bibinfo {author} {\bibfnamefont {Tobias}\ \bibnamefont
  {Micklitz}}, \bibinfo {author} {\bibfnamefont {J\'er\^ome}\ \bibnamefont
  {Rech}}, \ and\ \bibinfo {author} {\bibfnamefont {K.~A.}\ \bibnamefont
  {Matveev}},\ }\bibfield  {title} {\enquote {\bibinfo {title} {Transport
  properties of partially equilibrated quantum wires},}\ }\href {\doibase
  10.1103/PhysRevB.81.115313} {\bibfield  {journal} {\bibinfo  {journal} {Phys.
  Rev. B}\ }\textbf {\bibinfo {volume} {81}},\ \bibinfo {pages} {115313}
  (\bibinfo {year} {2010})}\BibitemShut {NoStop}%
\bibitem [{\citenamefont {Levchenko}\ \emph {et~al.}(2011)\citenamefont
  {Levchenko}, \citenamefont {Micklitz}, \citenamefont {Ristivojevic},\ and\
  \citenamefont {Matveev}}]{Levchenko2011}%
  \BibitemOpen
  \bibfield  {author} {\bibinfo {author} {\bibfnamefont {Alex}\ \bibnamefont
  {Levchenko}}, \bibinfo {author} {\bibfnamefont {Tobias}\ \bibnamefont
  {Micklitz}}, \bibinfo {author} {\bibfnamefont {Zoran}\ \bibnamefont
  {Ristivojevic}}, \ and\ \bibinfo {author} {\bibfnamefont {K.~A.}\
  \bibnamefont {Matveev}},\ }\bibfield  {title} {\enquote {\bibinfo {title}
  {Interaction effects on thermal transport in quantum wires},}\ }\href
  {\doibase 10.1103/PhysRevB.84.115447} {\bibfield  {journal} {\bibinfo
  {journal} {Phys. Rev. B}\ }\textbf {\bibinfo {volume} {84}},\ \bibinfo
  {pages} {115447} (\bibinfo {year} {2011})}\BibitemShut {NoStop}%
\bibitem [{\citenamefont {Teber}(2012)}]{Teber2012}%
  \BibitemOpen
  \bibfield  {author} {\bibinfo {author} {\bibfnamefont {S.}~\bibnamefont
  {Teber}},\ }\bibfield  {title} {\enquote {\bibinfo {title} {One-dimensional
  interacting electrons beyond the {Dzyaloshinskii-Larkin} theorem},}\ }\href
  {\doibase 10.1103/PhysRevB.86.195112} {\bibfield  {journal} {\bibinfo
  {journal} {Phys. Rev. B}\ }\textbf {\bibinfo {volume} {86}},\ \bibinfo
  {pages} {195112} (\bibinfo {year} {2012})}\BibitemShut {NoStop}%
\bibitem [{\citenamefont {Ristivojevic}\ and\ \citenamefont
  {Matveev}(2013)}]{Ristivojevic2013}%
  \BibitemOpen
  \bibfield  {author} {\bibinfo {author} {\bibfnamefont {Zoran}\ \bibnamefont
  {Ristivojevic}}\ and\ \bibinfo {author} {\bibfnamefont {K.~A.}\ \bibnamefont
  {Matveev}},\ }\bibfield  {title} {\enquote {\bibinfo {title} {Relaxation of
  weakly interacting electrons in one dimension},}\ }\href {\doibase
  10.1103/PhysRevB.87.165108} {\bibfield  {journal} {\bibinfo  {journal} {Phys.
  Rev. B}\ }\textbf {\bibinfo {volume} {87}},\ \bibinfo {pages} {165108}
  (\bibinfo {year} {2013})}\BibitemShut {NoStop}%
\bibitem [{\citenamefont {Protopopov}\ \emph {et~al.}(2014)\citenamefont
  {Protopopov}, \citenamefont {Gutman},\ and\ \citenamefont
  {Mirlin}}]{Protopopov2014}%
  \BibitemOpen
  \bibfield  {author} {\bibinfo {author} {\bibfnamefont {I.~V.}\ \bibnamefont
  {Protopopov}}, \bibinfo {author} {\bibfnamefont {D.~B.}\ \bibnamefont
  {Gutman}}, \ and\ \bibinfo {author} {\bibfnamefont {A.~D.}\ \bibnamefont
  {Mirlin}},\ }\bibfield  {title} {\enquote {\bibinfo {title} {Relaxation in
  {Luttinger} liquids: {Bose-Fermi} duality},}\ }\href {\doibase
  10.1103/PhysRevB.90.125113} {\bibfield  {journal} {\bibinfo  {journal} {Phys.
  Rev. B}\ }\textbf {\bibinfo {volume} {90}},\ \bibinfo {pages} {125113}
  (\bibinfo {year} {2014})}\BibitemShut {NoStop}%
\bibitem [{\citenamefont {Rieder}\ \emph {et~al.}(2014)\citenamefont {Rieder},
  \citenamefont {Micklitz}, \citenamefont {Levchenko},\ and\ \citenamefont
  {Matveev}}]{Rieder2014}%
  \BibitemOpen
  \bibfield  {author} {\bibinfo {author} {\bibfnamefont {M.-T.}\ \bibnamefont
  {Rieder}}, \bibinfo {author} {\bibfnamefont {T.}~\bibnamefont {Micklitz}},
  \bibinfo {author} {\bibfnamefont {A.}~\bibnamefont {Levchenko}}, \ and\
  \bibinfo {author} {\bibfnamefont {K.~A.}\ \bibnamefont {Matveev}},\
  }\bibfield  {title} {\enquote {\bibinfo {title} {Interaction-induced
  backscattering in short quantum wires},}\ }\href {\doibase
  10.1103/PhysRevB.90.165405} {\bibfield  {journal} {\bibinfo  {journal} {Phys.
  Rev. B}\ }\textbf {\bibinfo {volume} {90}},\ \bibinfo {pages} {165405}
  (\bibinfo {year} {2014})}\BibitemShut {NoStop}%
\bibitem [{\citenamefont {Protopopov}\ \emph {et~al.}(2015)\citenamefont
  {Protopopov}, \citenamefont {Gutman},\ and\ \citenamefont
  {Mirlin}}]{Protopopov2015}%
  \BibitemOpen
  \bibfield  {author} {\bibinfo {author} {\bibfnamefont {I.~V.}\ \bibnamefont
  {Protopopov}}, \bibinfo {author} {\bibfnamefont {D.~B.}\ \bibnamefont
  {Gutman}}, \ and\ \bibinfo {author} {\bibfnamefont {A.~D.}\ \bibnamefont
  {Mirlin}},\ }\bibfield  {title} {\enquote {\bibinfo {title} {Equilibration in
  a chiral {Luttinger} liquid},}\ }\href {\doibase 10.1103/PhysRevB.91.195110}
  {\bibfield  {journal} {\bibinfo  {journal} {Phys. Rev. B}\ }\textbf {\bibinfo
  {volume} {91}},\ \bibinfo {pages} {195110} (\bibinfo {year}
  {2015})}\BibitemShut {NoStop}%
\bibitem [{\citenamefont {Ristivojevic}\ and\ \citenamefont
  {Matveev}(2021)}]{Ristivojevic2021}%
  \BibitemOpen
  \bibfield  {author} {\bibinfo {author} {\bibfnamefont {Zoran}\ \bibnamefont
  {Ristivojevic}}\ and\ \bibinfo {author} {\bibfnamefont {K.~A.}\ \bibnamefont
  {Matveev}},\ }\bibfield  {title} {\enquote {\bibinfo {title} {Quasiparticle
  energy relaxation in a gas of one-dimensional fermions with {Coulomb}
  interaction},}\ }\href {\doibase 10.1103/PhysRevLett.127.086803} {\bibfield
  {journal} {\bibinfo  {journal} {Phys. Rev. Lett.}\ }\textbf {\bibinfo
  {volume} {127}},\ \bibinfo {pages} {086803} (\bibinfo {year}
  {2021})}\BibitemShut {NoStop}%
\bibitem [{\citenamefont {Apostolov}\ \emph {et~al.}(2014)\citenamefont
  {Apostolov}, \citenamefont {Levchenko},\ and\ \citenamefont
  {Andreev}}]{ALA2014}%
  \BibitemOpen
  \bibfield  {author} {\bibinfo {author} {\bibfnamefont {S.~S.}\ \bibnamefont
  {Apostolov}}, \bibinfo {author} {\bibfnamefont {A.}~\bibnamefont
  {Levchenko}}, \ and\ \bibinfo {author} {\bibfnamefont {A.~V.}\ \bibnamefont
  {Andreev}},\ }\bibfield  {title} {\enquote {\bibinfo {title} {Hydrodynamic
  {C}oulomb drag of strongly correlated electron liquids},}\ }\href@noop {}
  {\bibfield  {journal} {\bibinfo  {journal} {Phys. Rev. B}\ }\textbf {\bibinfo
  {volume} {89}},\ \bibinfo {pages} {121104(R)} (\bibinfo {year}
  {2014})}\BibitemShut {NoStop}%
\bibitem [{\citenamefont {Patel}\ \emph {et~al.}(2017)\citenamefont {Patel},
  \citenamefont {Davison},\ and\ \citenamefont {Levchenko}}]{Patel2017}%
  \BibitemOpen
  \bibfield  {author} {\bibinfo {author} {\bibfnamefont {Aavishkar~A.}\
  \bibnamefont {Patel}}, \bibinfo {author} {\bibfnamefont {Richard~A.}\
  \bibnamefont {Davison}}, \ and\ \bibinfo {author} {\bibfnamefont {Alex}\
  \bibnamefont {Levchenko}},\ }\bibfield  {title} {\enquote {\bibinfo {title}
  {Hydrodynamic flows of non-{Fermi} liquids: Magnetotransport and bilayer
  drag},}\ }\href {\doibase 10.1103/PhysRevB.96.205417} {\bibfield  {journal}
  {\bibinfo  {journal} {Phys. Rev. B}\ }\textbf {\bibinfo {volume} {96}},\
  \bibinfo {pages} {205417} (\bibinfo {year} {2017})}\BibitemShut {NoStop}%
\bibitem [{\citenamefont {Levchenko}\ \emph {et~al.}(2022)\citenamefont
  {Levchenko}, \citenamefont {Li},\ and\ \citenamefont {Andreev}}]{LLA2022}%
  \BibitemOpen
  \bibfield  {author} {\bibinfo {author} {\bibfnamefont {A.}~\bibnamefont
  {Levchenko}}, \bibinfo {author} {\bibfnamefont {S.}~\bibnamefont {Li}}, \
  and\ \bibinfo {author} {\bibfnamefont {A.~V.}\ \bibnamefont {Andreev}},\
  }\bibfield  {title} {\enquote {\bibinfo {title} {Fluctuation-driven thermal
  transport in graphene double-layers at charge neutrality},}\ }\href@noop {}
  {\bibfield  {journal} {\bibinfo  {journal} {Phys. Rev. B}\ }\textbf {\bibinfo
  {volume} {106}},\ \bibinfo {pages} {125304} (\bibinfo {year}
  {2022})}\BibitemShut {NoStop}%
\bibitem [{\citenamefont {Zverevich}\ and\ \citenamefont
  {Levchenko}(2023)}]{ZL2023}%
  \BibitemOpen
  \bibfield  {author} {\bibinfo {author} {\bibfnamefont {D.}~\bibnamefont
  {Zverevich}}\ and\ \bibinfo {author} {\bibfnamefont {A.}~\bibnamefont
  {Levchenko}},\ }\bibfield  {title} {\enquote {\bibinfo {title} {Transport
  signatures of plasmon fluctuations in electron hydrodynamics},}\ }\href@noop
  {} {\bibfield  {journal} {\bibinfo  {journal} {Low Temp. Phys.}\ }\textbf
  {\bibinfo {volume} {49}},\ \bibinfo {pages} {1376} (\bibinfo {year}
  {2023})}\BibitemShut {NoStop}%
\bibitem [{\citenamefont {Levchenko}\ \emph {et~al.}(2026)\citenamefont
  {Levchenko}, \citenamefont {Kirkinis},\ and\ \citenamefont
  {Andreev}}]{Levchenko2026}%
  \BibitemOpen
  \bibfield  {author} {\bibinfo {author} {\bibfnamefont {A.}~\bibnamefont
  {Levchenko}}, \bibinfo {author} {\bibfnamefont {E.}~\bibnamefont {Kirkinis}},
  \ and\ \bibinfo {author} {\bibfnamefont {A.~V.}\ \bibnamefont {Andreev}},\
  }\bibfield  {title} {\enquote {\bibinfo {title} {Fluctuation-induced giant
  magnetoresistance in charge-neutral graphene},}\ }\href {\doibase
  10.1103/12ft-16kb} {\bibfield  {journal} {\bibinfo  {journal} {Phys. Rev.
  Lett.}\ }\textbf {\bibinfo {volume} {136}},\ \bibinfo {pages} {126301}
  (\bibinfo {year} {2026})}\BibitemShut {NoStop}%
\bibitem [{\citenamefont {Matveev}\ and\ \citenamefont
  {Andreev}(2019)}]{MA2019}%
  \BibitemOpen
  \bibfield  {author} {\bibinfo {author} {\bibfnamefont {K.~A.}\ \bibnamefont
  {Matveev}}\ and\ \bibinfo {author} {\bibfnamefont {A.~V.}\ \bibnamefont
  {Andreev}},\ }\bibfield  {title} {\enquote {\bibinfo {title} {Two-fluid
  dynamics of one-dimensional quantum liquids in the absence of {G}alilean
  invariance},}\ }\href@noop {} {\bibfield  {journal} {\bibinfo  {journal}
  {Phys. Rev. B}\ }\textbf {\bibinfo {volume} {100}},\ \bibinfo {pages}
  {035418} (\bibinfo {year} {2019})}\BibitemShut {NoStop}%
\bibitem [{\citenamefont {Protopopov}\ \emph {et~al.}(2021)\citenamefont
  {Protopopov}, \citenamefont {Samanta}, \citenamefont {Mirlin},\ and\
  \citenamefont {Gutman}}]{PSMG2021}%
  \BibitemOpen
  \bibfield  {author} {\bibinfo {author} {\bibfnamefont {I.~V.}\ \bibnamefont
  {Protopopov}}, \bibinfo {author} {\bibfnamefont {R.}~\bibnamefont {Samanta}},
  \bibinfo {author} {\bibfnamefont {A.~D.}\ \bibnamefont {Mirlin}}, \ and\
  \bibinfo {author} {\bibfnamefont {D.~B.}\ \bibnamefont {Gutman}},\ }\bibfield
   {title} {\enquote {\bibinfo {title} {Anomalous hydrodynamics in a
  one-dimensional electronic fluid},}\ }\href@noop {} {\bibfield  {journal}
  {\bibinfo  {journal} {Phys. Rev. Lett.}\ }\textbf {\bibinfo {volume} {126}},\
  \bibinfo {pages} {256801} (\bibinfo {year} {2021})}\BibitemShut {NoStop}%
\bibitem [{\citenamefont {Matveev}\ and\ \citenamefont
  {Andreev}(2017)}]{MA2017}%
  \BibitemOpen
  \bibfield  {author} {\bibinfo {author} {\bibfnamefont {K.~A.}\ \bibnamefont
  {Matveev}}\ and\ \bibinfo {author} {\bibfnamefont {A.~V.}\ \bibnamefont
  {Andreev}},\ }\bibfield  {title} {\enquote {\bibinfo {title} {Second sound in
  systems of one-dimensional fermions},}\ }\href@noop {} {\bibfield  {journal}
  {\bibinfo  {journal} {Phys. Rev. Lett.}\ }\textbf {\bibinfo {volume} {119}},\
  \bibinfo {pages} {266801} (\bibinfo {year} {2017})}\BibitemShut {NoStop}%
\bibitem [{\citenamefont {Matveev}\ and\ \citenamefont
  {Andreev}(2018{\natexlab{a}})}]{MA2018}%
  \BibitemOpen
  \bibfield  {author} {\bibinfo {author} {\bibfnamefont {K.~A.}\ \bibnamefont
  {Matveev}}\ and\ \bibinfo {author} {\bibfnamefont {A.~V.}\ \bibnamefont
  {Andreev}},\ }\bibfield  {title} {\enquote {\bibinfo {title} {Propagation and
  attenuation of sound in one-dimensional quantum liquids},}\ }\href {\doibase
  10.1103/PhysRevB.98.155441} {\bibfield  {journal} {\bibinfo  {journal} {Phys.
  Rev. B}\ }\textbf {\bibinfo {volume} {98}},\ \bibinfo {pages} {155441}
  (\bibinfo {year} {2018}{\natexlab{a}})}\BibitemShut {NoStop}%
\bibitem [{\citenamefont {Matveev}\ and\ \citenamefont
  {Andreev}(2018{\natexlab{b}})}]{MAhybrid2018}%
  \BibitemOpen
  \bibfield  {author} {\bibinfo {author} {\bibfnamefont {K.~A.}\ \bibnamefont
  {Matveev}}\ and\ \bibinfo {author} {\bibfnamefont {A.~V.}\ \bibnamefont
  {Andreev}},\ }\bibfield  {title} {\enquote {\bibinfo {title} {Hybrid sound
  modes in one-dimensional quantum liquids},}\ }\href {\doibase
  10.1103/PhysRevLett.121.026803} {\bibfield  {journal} {\bibinfo  {journal}
  {Phys. Rev. Lett.}\ }\textbf {\bibinfo {volume} {121}},\ \bibinfo {pages}
  {026803} (\bibinfo {year} {2018}{\natexlab{b}})}\BibitemShut {NoStop}%
\bibitem [{\citenamefont {Matveev}\ and\ \citenamefont
  {Ristivojevic}(2019)}]{MR2015}%
  \BibitemOpen
  \bibfield  {author} {\bibinfo {author} {\bibfnamefont {K.~A.}\ \bibnamefont
  {Matveev}}\ and\ \bibinfo {author} {\bibfnamefont {Zoran}\ \bibnamefont
  {Ristivojevic}},\ }\bibfield  {title} {\enquote {\bibinfo {title} {Thermal
  conductivity of the degenerate one-dimensional {Fermi} gas},}\ }\href
  {\doibase 10.1103/PhysRevB.99.155428} {\bibfield  {journal} {\bibinfo
  {journal} {Phys. Rev. B}\ }\textbf {\bibinfo {volume} {99}},\ \bibinfo
  {pages} {155428} (\bibinfo {year} {2019})}\BibitemShut {NoStop}%
\bibitem [{\citenamefont {Samanta}\ \emph {et~al.}(2019)\citenamefont
  {Samanta}, \citenamefont {Protopopov}, \citenamefont {Mirlin},\ and\
  \citenamefont {Gutman}}]{SPMG2019}%
  \BibitemOpen
  \bibfield  {author} {\bibinfo {author} {\bibfnamefont {R.}~\bibnamefont
  {Samanta}}, \bibinfo {author} {\bibfnamefont {I.~V.}\ \bibnamefont
  {Protopopov}}, \bibinfo {author} {\bibfnamefont {A.~D.}\ \bibnamefont
  {Mirlin}}, \ and\ \bibinfo {author} {\bibfnamefont {D.~B.}\ \bibnamefont
  {Gutman}},\ }\bibfield  {title} {\enquote {\bibinfo {title} {Thermal
  transport in one-dimensional electronic fluid},}\ }\href@noop {} {\bibfield
  {journal} {\bibinfo  {journal} {Phys. Rev. Lett.}\ }\textbf {\bibinfo
  {volume} {122}},\ \bibinfo {pages} {206801} (\bibinfo {year}
  {2019})}\BibitemShut {NoStop}%
\bibitem [{\citenamefont {Gutman}\ \emph {et~al.}(2023)\citenamefont {Gutman},
  \citenamefont {Protopopov}, \citenamefont {Samanta},\ and\ \citenamefont
  {Mirlin}}]{GPSM2023}%
  \BibitemOpen
  \bibfield  {author} {\bibinfo {author} {\bibfnamefont {D.~B.}\ \bibnamefont
  {Gutman}}, \bibinfo {author} {\bibfnamefont {I.~V.}\ \bibnamefont
  {Protopopov}}, \bibinfo {author} {\bibfnamefont {R.}~\bibnamefont {Samanta}},
  \ and\ \bibinfo {author} {\bibfnamefont {A.~D.}\ \bibnamefont {Mirlin}},\
  }\bibfield  {title} {\enquote {\bibinfo {title} {Thermal conductivity in
  one-dimensional electronic fluids},}\ }\href@noop {} {\bibfield  {journal}
  {\bibinfo  {journal} {Low Temp. Phys.}\ }\textbf {\bibinfo {volume} {49}},\
  \bibinfo {pages} {1358} (\bibinfo {year} {2023})}\BibitemShut {NoStop}%
\bibitem [{\citenamefont {DeGottardi}\ and\ \citenamefont
  {Matveev}(2018)}]{DGM2018}%
  \BibitemOpen
  \bibfield  {author} {\bibinfo {author} {\bibfnamefont {W.}~\bibnamefont
  {DeGottardi}}\ and\ \bibinfo {author} {\bibfnamefont {K.~A.}\ \bibnamefont
  {Matveev}},\ }\bibfield  {title} {\enquote {\bibinfo {title} {Viscosity of a
  multichannel one-dimensional {F}ermi gas},}\ }\href@noop {} {\bibfield
  {journal} {\bibinfo  {journal} {Phys. Rev. B}\ }\textbf {\bibinfo {volume}
  {97}},\ \bibinfo {pages} {045135} (\bibinfo {year} {2018})}\BibitemShut
  {NoStop}%
\bibitem [{\citenamefont {Gramila}\ \emph {et~al.}(1991)\citenamefont
  {Gramila}, \citenamefont {Eisenstein}, \citenamefont {MacDonald},
  \citenamefont {Pfeiffer},\ and\ \citenamefont {West}}]{Gramila1991}%
  \BibitemOpen
  \bibfield  {author} {\bibinfo {author} {\bibfnamefont {T.~J.}\ \bibnamefont
  {Gramila}}, \bibinfo {author} {\bibfnamefont {J.~P.}\ \bibnamefont
  {Eisenstein}}, \bibinfo {author} {\bibfnamefont {A.~H.}\ \bibnamefont
  {MacDonald}}, \bibinfo {author} {\bibfnamefont {L.~N.}\ \bibnamefont
  {Pfeiffer}}, \ and\ \bibinfo {author} {\bibfnamefont {K.~W.}\ \bibnamefont
  {West}},\ }\bibfield  {title} {\enquote {\bibinfo {title} {Mutual friction
  between parallel two-dimensional electron systems},}\ }\href@noop {}
  {\bibfield  {journal} {\bibinfo  {journal} {Phys. Rev. Lett.}\ }\textbf
  {\bibinfo {volume} {66}},\ \bibinfo {pages} {1216} (\bibinfo {year}
  {1991})}\BibitemShut {NoStop}%
\bibitem [{\citenamefont {Narozhny}\ and\ \citenamefont
  {Levchenko}(2016)}]{NarozhnyLevchenko2016}%
  \BibitemOpen
  \bibfield  {author} {\bibinfo {author} {\bibfnamefont {B.~N.}\ \bibnamefont
  {Narozhny}}\ and\ \bibinfo {author} {\bibfnamefont {A.}~\bibnamefont
  {Levchenko}},\ }\bibfield  {title} {\enquote {\bibinfo {title} {Coulomb
  drag},}\ }\href@noop {} {\bibfield  {journal} {\bibinfo  {journal} {Rev. Mod.
  Phys.}\ }\textbf {\bibinfo {volume} {88}},\ \bibinfo {pages} {025003}
  (\bibinfo {year} {2016})}\BibitemShut {NoStop}%
\bibitem [{\citenamefont {Forster}(1975)}]{ForsterBook}%
  \BibitemOpen
  \bibfield  {author} {\bibinfo {author} {\bibfnamefont {D.}~\bibnamefont
  {Forster}},\ }\href@noop {} {\emph {\bibinfo {title} {Hydrodynamic
  Fluctuations, Broken Symmetry, and Correlation Functions}}}\ (\bibinfo
  {publisher} {Benjamin},\ \bibinfo {address} {Reading, MA},\ \bibinfo {year}
  {1975})\BibitemShut {NoStop}%
\bibitem [{\citenamefont {Pustilnik}\ \emph {et~al.}(2006)\citenamefont
  {Pustilnik}, \citenamefont {Khodas}, \citenamefont {Kamenev},\ and\
  \citenamefont {Glazman}}]{PKKG2006}%
  \BibitemOpen
  \bibfield  {author} {\bibinfo {author} {\bibfnamefont {M.}~\bibnamefont
  {Pustilnik}}, \bibinfo {author} {\bibfnamefont {M.}~\bibnamefont {Khodas}},
  \bibinfo {author} {\bibfnamefont {A.}~\bibnamefont {Kamenev}}, \ and\
  \bibinfo {author} {\bibfnamefont {L.~I.}\ \bibnamefont {Glazman}},\
  }\bibfield  {title} {\enquote {\bibinfo {title} {Dynamic response of
  one-dimensional interacting fermions},}\ }\href@noop {} {\bibfield  {journal}
  {\bibinfo  {journal} {Phys. Rev. Lett.}\ }\textbf {\bibinfo {volume} {96}},\
  \bibinfo {pages} {196405} (\bibinfo {year} {2006})}\BibitemShut {NoStop}%
\bibitem [{\citenamefont {Pogrebinskii}(1977)}]{Pogrebinskii1977}%
  \BibitemOpen
  \bibfield  {author} {\bibinfo {author} {\bibfnamefont {M.~B.}\ \bibnamefont
  {Pogrebinskii}},\ }\bibfield  {title} {\enquote {\bibinfo {title} {Mutual
  drag of carriers in a semiconductor-insulator-semiconductor system},}\
  }\href@noop {} {\bibfield  {journal} {\bibinfo  {journal} {Fiz. Tekh.
  Poluprovodn.}\ }\textbf {\bibinfo {volume} {11}},\ \bibinfo {pages} {637}
  (\bibinfo {year} {1977})},\ \bibinfo {note} {[Sov. Phys. Semicond. {\bf 11},
  372 (1977)]}\BibitemShut {NoStop}%
\bibitem [{\citenamefont {Price}(1983)}]{Price1983}%
  \BibitemOpen
  \bibfield  {author} {\bibinfo {author} {\bibfnamefont {P.~J.}\ \bibnamefont
  {Price}},\ }\bibfield  {title} {\enquote {\bibinfo {title} {Hot electron
  effects in heterolayers},}\ }\href@noop {} {\bibfield  {journal} {\bibinfo
  {journal} {Physica B+C}\ }\textbf {\bibinfo {volume} {117}},\ \bibinfo
  {pages} {750} (\bibinfo {year} {1983})}\BibitemShut {NoStop}%
\bibitem [{\citenamefont {Pustilnik}\ \emph {et~al.}(2003)\citenamefont
  {Pustilnik}, \citenamefont {Mishchenko}, \citenamefont {Glazman},\ and\
  \citenamefont {Andreev}}]{PMGA2003}%
  \BibitemOpen
  \bibfield  {author} {\bibinfo {author} {\bibfnamefont {M.}~\bibnamefont
  {Pustilnik}}, \bibinfo {author} {\bibfnamefont {E.~G.}\ \bibnamefont
  {Mishchenko}}, \bibinfo {author} {\bibfnamefont {L.~I.}\ \bibnamefont
  {Glazman}}, \ and\ \bibinfo {author} {\bibfnamefont {A.~V.}\ \bibnamefont
  {Andreev}},\ }\bibfield  {title} {\enquote {\bibinfo {title} {Coulomb drag by
  small momentum transfer between quantum wires},}\ }\href@noop {} {\bibfield
  {journal} {\bibinfo  {journal} {Phys. Rev. Lett.}\ }\textbf {\bibinfo
  {volume} {91}},\ \bibinfo {pages} {126805} (\bibinfo {year}
  {2003})}\BibitemShut {NoStop}%
\bibitem [{\citenamefont {Dmitriev}\ \emph {et~al.}(2012)\citenamefont
  {Dmitriev}, \citenamefont {Gornyi},\ and\ \citenamefont
  {Polyakov}}]{DGP2012}%
  \BibitemOpen
  \bibfield  {author} {\bibinfo {author} {\bibfnamefont {A.~P.}\ \bibnamefont
  {Dmitriev}}, \bibinfo {author} {\bibfnamefont {I.~V.}\ \bibnamefont
  {Gornyi}}, \ and\ \bibinfo {author} {\bibfnamefont {D.~G.}\ \bibnamefont
  {Polyakov}},\ }\bibfield  {title} {\enquote {\bibinfo {title} {Coulomb drag
  between ballistic quantum wires},}\ }\href@noop {} {\bibfield  {journal}
  {\bibinfo  {journal} {Phys. Rev. B}\ }\textbf {\bibinfo {volume} {86}},\
  \bibinfo {pages} {245402} (\bibinfo {year} {2012})}\BibitemShut {NoStop}%
\bibitem [{\citenamefont {Klesse}\ and\ \citenamefont
  {Stern}(2000)}]{KlesseStern2000}%
  \BibitemOpen
  \bibfield  {author} {\bibinfo {author} {\bibfnamefont {R.}~\bibnamefont
  {Klesse}}\ and\ \bibinfo {author} {\bibfnamefont {A.}~\bibnamefont {Stern}},\
  }\bibfield  {title} {\enquote {\bibinfo {title} {Coulomb drag between quantum
  wires},}\ }\href@noop {} {\bibfield  {journal} {\bibinfo  {journal} {Phys.
  Rev. B}\ }\textbf {\bibinfo {volume} {62}},\ \bibinfo {pages} {16912}
  (\bibinfo {year} {2000})}\BibitemShut {NoStop}%
\bibitem [{\citenamefont {Nazarov}\ and\ \citenamefont
  {Averin}(1998)}]{NazarovAverin1998}%
  \BibitemOpen
  \bibfield  {author} {\bibinfo {author} {\bibfnamefont {Yu.~V.}\ \bibnamefont
  {Nazarov}}\ and\ \bibinfo {author} {\bibfnamefont {D.~V.}\ \bibnamefont
  {Averin}},\ }\bibfield  {title} {\enquote {\bibinfo {title} {Current drag in
  capacitively coupled {L}uttinger constrictions},}\ }\href@noop {} {\bibfield
  {journal} {\bibinfo  {journal} {Phys. Rev. Lett.}\ }\textbf {\bibinfo
  {volume} {81}},\ \bibinfo {pages} {653} (\bibinfo {year} {1998})}\BibitemShut
  {NoStop}%
\bibitem [{\citenamefont {Laroche}\ \emph {et~al.}(2011)\citenamefont
  {Laroche}, \citenamefont {Gervais}, \citenamefont {Lilly},\ and\
  \citenamefont {Reno}}]{Laroche2011}%
  \BibitemOpen
  \bibfield  {author} {\bibinfo {author} {\bibfnamefont {D.}~\bibnamefont
  {Laroche}}, \bibinfo {author} {\bibfnamefont {G.}~\bibnamefont {Gervais}},
  \bibinfo {author} {\bibfnamefont {M.~P.}\ \bibnamefont {Lilly}}, \ and\
  \bibinfo {author} {\bibfnamefont {J.~L.}\ \bibnamefont {Reno}},\ }\bibfield
  {title} {\enquote {\bibinfo {title} {Positive and negative {C}oulomb drag in
  vertically integrated one-dimensional quantum wires},}\ }\href@noop {}
  {\bibfield  {journal} {\bibinfo  {journal} {Nat. Nanotechnol.}\ }\textbf
  {\bibinfo {volume} {6}},\ \bibinfo {pages} {793} (\bibinfo {year}
  {2011})}\BibitemShut {NoStop}%
\bibitem [{\citenamefont {Laroche}\ \emph {et~al.}(2014)\citenamefont
  {Laroche}, \citenamefont {Gervais}, \citenamefont {Lilly},\ and\
  \citenamefont {Reno}}]{Laroche2014}%
  \BibitemOpen
  \bibfield  {author} {\bibinfo {author} {\bibfnamefont {D.}~\bibnamefont
  {Laroche}}, \bibinfo {author} {\bibfnamefont {G.}~\bibnamefont {Gervais}},
  \bibinfo {author} {\bibfnamefont {M.~P.}\ \bibnamefont {Lilly}}, \ and\
  \bibinfo {author} {\bibfnamefont {J.~L.}\ \bibnamefont {Reno}},\ }\bibfield
  {title} {\enquote {\bibinfo {title} {1{D}-1{D} {C}oulomb drag signature of a
  {L}uttinger liquid},}\ }\href@noop {} {\bibfield  {journal} {\bibinfo
  {journal} {Science}\ }\textbf {\bibinfo {volume} {343}},\ \bibinfo {pages}
  {631} (\bibinfo {year} {2014})}\BibitemShut {NoStop}%
\bibitem [{\citenamefont {Zheng}\ \emph
  {et~al.}(2025{\natexlab{a}})\citenamefont {Zheng}, \citenamefont {Makaju},
  \citenamefont {Gazizulin}, \citenamefont {Addamane},\ and\ \citenamefont
  {Laroche}}]{Makaju2025}%
  \BibitemOpen
  \bibfield  {author} {\bibinfo {author} {\bibfnamefont {M.}~\bibnamefont
  {Zheng}}, \bibinfo {author} {\bibfnamefont {R.}~\bibnamefont {Makaju}},
  \bibinfo {author} {\bibfnamefont {R.}~\bibnamefont {Gazizulin}}, \bibinfo
  {author} {\bibfnamefont {S.~J.}\ \bibnamefont {Addamane}}, \ and\ \bibinfo
  {author} {\bibfnamefont {D.}~\bibnamefont {Laroche}},\ }\bibfield  {title}
  {\enquote {\bibinfo {title} {Tunable reciprocal and nonreciprocal
  contributions to {1D} {Coulomb} drag},}\ }\href {\doibase
  10.1038/s41467-025-62324-6} {\bibfield  {journal} {\bibinfo  {journal} {Nat.
  Commun.}\ }\textbf {\bibinfo {volume} {16}},\ \bibinfo {pages} {6963}
  (\bibinfo {year} {2025}{\natexlab{a}})}\BibitemShut {NoStop}%
\bibitem [{\citenamefont {Zheng}\ \emph
  {et~al.}(2025{\natexlab{b}})\citenamefont {Zheng}, \citenamefont {Makaju},
  \citenamefont {Gazizulin}, \citenamefont {Levchenko}, \citenamefont
  {Addamane},\ and\ \citenamefont {Laroche}}]{Zheng2025}%
  \BibitemOpen
  \bibfield  {author} {\bibinfo {author} {\bibfnamefont {Mingyang}\
  \bibnamefont {Zheng}}, \bibinfo {author} {\bibfnamefont {Rebika}\
  \bibnamefont {Makaju}}, \bibinfo {author} {\bibfnamefont {Rasul}\
  \bibnamefont {Gazizulin}}, \bibinfo {author} {\bibfnamefont {Alex}\
  \bibnamefont {Levchenko}}, \bibinfo {author} {\bibfnamefont {Sadhvikas~J.}\
  \bibnamefont {Addamane}}, \ and\ \bibinfo {author} {\bibfnamefont
  {Dominique}\ \bibnamefont {Laroche}},\ }\bibfield  {title} {\enquote
  {\bibinfo {title} {Quasi-{1D} {Coulomb} drag in the nonlinear regime},}\
  }\href {\doibase 10.1103/v3dn-bnrp} {\bibfield  {journal} {\bibinfo
  {journal} {Phys. Rev. Lett.}\ }\textbf {\bibinfo {volume} {134}},\ \bibinfo
  {pages} {236301} (\bibinfo {year} {2025}{\natexlab{b}})}\BibitemShut
  {NoStop}%
\bibitem [{\citenamefont {Zheng}\ \emph {et~al.}(2026)\citenamefont {Zheng},
  \citenamefont {Makaju}, \citenamefont {Gazizulin}, \citenamefont {Levchenko},
  \citenamefont {Addamane},\ and\ \citenamefont {Laroche}}]{Zheng2026}%
  \BibitemOpen
  \bibfield  {author} {\bibinfo {author} {\bibfnamefont {Mingyang}\
  \bibnamefont {Zheng}}, \bibinfo {author} {\bibfnamefont {Rebika}\
  \bibnamefont {Makaju}}, \bibinfo {author} {\bibfnamefont {Rasul}\
  \bibnamefont {Gazizulin}}, \bibinfo {author} {\bibfnamefont {Alex}\
  \bibnamefont {Levchenko}}, \bibinfo {author} {\bibfnamefont {Sadhvikas~J.}\
  \bibnamefont {Addamane}}, \ and\ \bibinfo {author} {\bibfnamefont
  {Dominique}\ \bibnamefont {Laroche}},\ }\bibfield  {title} {\enquote
  {\bibinfo {title} {Quasi-one-dimensional {Coulomb} drag between
  spin-polarized quantum wires},}\ }\href {\doibase 10.1103/5yqq-cw73}
  {\bibfield  {journal} {\bibinfo  {journal} {Phys. Rev. B}\ }\textbf {\bibinfo
  {volume} {113}},\ \bibinfo {pages} {L121408} (\bibinfo {year}
  {2026})}\BibitemShut {NoStop}%
\bibitem [{\citenamefont {Polder}\ and\ \citenamefont
  {Van~Hove}(1971)}]{PolderVanHove1971}%
  \BibitemOpen
  \bibfield  {author} {\bibinfo {author} {\bibfnamefont {D.}~\bibnamefont
  {Polder}}\ and\ \bibinfo {author} {\bibfnamefont {M.}~\bibnamefont
  {Van~Hove}},\ }\bibfield  {title} {\enquote {\bibinfo {title} {Theory of
  radiative heat transfer between closely spaced bodies},}\ }\href@noop {}
  {\bibfield  {journal} {\bibinfo  {journal} {Phys. Rev. B}\ }\textbf {\bibinfo
  {volume} {4}},\ \bibinfo {pages} {3303} (\bibinfo {year} {1971})}\BibitemShut
  {NoStop}%
\bibitem [{\citenamefont {Volokitin}\ and\ \citenamefont
  {Persson}(2007)}]{VolokitinPersson2007}%
  \BibitemOpen
  \bibfield  {author} {\bibinfo {author} {\bibfnamefont {A.~I.}\ \bibnamefont
  {Volokitin}}\ and\ \bibinfo {author} {\bibfnamefont {B.~N.~J.}\ \bibnamefont
  {Persson}},\ }\bibfield  {title} {\enquote {\bibinfo {title} {Near-field
  radiative heat transfer and noncontact friction},}\ }\href@noop {} {\bibfield
   {journal} {\bibinfo  {journal} {Rev. Mod. Phys.}\ }\textbf {\bibinfo
  {volume} {79}},\ \bibinfo {pages} {1291} (\bibinfo {year}
  {2007})}\BibitemShut {NoStop}%
\bibitem [{\citenamefont {Mahan}(2017)}]{Mahan2017}%
  \BibitemOpen
  \bibfield  {author} {\bibinfo {author} {\bibfnamefont {G.~D.}\ \bibnamefont
  {Mahan}},\ }\bibfield  {title} {\enquote {\bibinfo {title} {Tunneling of heat
  between metals},}\ }\href {\doibase 10.1103/PhysRevB.95.115427} {\bibfield
  {journal} {\bibinfo  {journal} {Phys. Rev. B}\ }\textbf {\bibinfo {volume}
  {95}},\ \bibinfo {pages} {115427} (\bibinfo {year} {2017})}\BibitemShut
  {NoStop}%
\bibitem [{\citenamefont {Ying}\ and\ \citenamefont
  {Kamenev}(2020)}]{Kamenev2020}%
  \BibitemOpen
  \bibfield  {author} {\bibinfo {author} {\bibfnamefont {Xuzhe}\ \bibnamefont
  {Ying}}\ and\ \bibinfo {author} {\bibfnamefont {Alex}\ \bibnamefont
  {Kamenev}},\ }\bibfield  {title} {\enquote {\bibinfo {title} {Plasmonic
  tuning of near-field heat transfer between graphene monolayers},}\ }\href
  {\doibase 10.1103/PhysRevB.102.195426} {\bibfield  {journal} {\bibinfo
  {journal} {Phys. Rev. B}\ }\textbf {\bibinfo {volume} {102}},\ \bibinfo
  {pages} {195426} (\bibinfo {year} {2020})}\BibitemShut {NoStop}%
\bibitem [{\citenamefont {Wise}\ and\ \citenamefont {Basko}(2021)}]{Basko2021}%
  \BibitemOpen
  \bibfield  {author} {\bibinfo {author} {\bibfnamefont {Jonathan~L.}\
  \bibnamefont {Wise}}\ and\ \bibinfo {author} {\bibfnamefont {Denis~M.}\
  \bibnamefont {Basko}},\ }\bibfield  {title} {\enquote {\bibinfo {title} {Near
  field versus far field in radiative heat transfer between two-dimensional
  metals},}\ }\href {\doibase 10.1103/PhysRevB.103.165423} {\bibfield
  {journal} {\bibinfo  {journal} {Phys. Rev. B}\ }\textbf {\bibinfo {volume}
  {103}},\ \bibinfo {pages} {165423} (\bibinfo {year} {2021})}\BibitemShut
  {NoStop}%
\bibitem [{\citenamefont {Bhandari}\ \emph {et~al.}(2018)\citenamefont
  {Bhandari}, \citenamefont {Chiriac{\`o}}, \citenamefont {Erdman},
  \citenamefont {Fazio},\ and\ \citenamefont {Taddei}}]{Bhandari2018}%
  \BibitemOpen
  \bibfield  {author} {\bibinfo {author} {\bibfnamefont {B.}~\bibnamefont
  {Bhandari}}, \bibinfo {author} {\bibfnamefont {G.}~\bibnamefont
  {Chiriac{\`o}}}, \bibinfo {author} {\bibfnamefont {P.~A.}\ \bibnamefont
  {Erdman}}, \bibinfo {author} {\bibfnamefont {R.}~\bibnamefont {Fazio}}, \
  and\ \bibinfo {author} {\bibfnamefont {F.}~\bibnamefont {Taddei}},\
  }\bibfield  {title} {\enquote {\bibinfo {title} {Thermal drag in electronic
  conductors},}\ }\href@noop {} {\bibfield  {journal} {\bibinfo  {journal}
  {Phys. Rev. B}\ }\textbf {\bibinfo {volume} {98}},\ \bibinfo {pages} {035415}
  (\bibinfo {year} {2018})}\BibitemShut {NoStop}%
\bibitem [{\citenamefont {Andreev}(1978)}]{Andreev1978}%
  \BibitemOpen
  \bibfield  {author} {\bibinfo {author} {\bibfnamefont {A.~F.}\ \bibnamefont
  {Andreev}},\ }\bibfield  {title} {\enquote {\bibinfo {title} {Corrections to
  the hydrodynamics of liquids},}\ }\href@noop {} {\bibfield  {journal}
  {\bibinfo  {journal} {Zh. Eksp. Teor. Fiz.}\ }\textbf {\bibinfo {volume}
  {75}},\ \bibinfo {pages} {1132} (\bibinfo {year} {1978})},\ \bibinfo {note}
  {[Sov. Phys. JETP {\bf 48}, 570 (1978)]}\BibitemShut {NoStop}%
\bibitem [{\citenamefont {Andreev}(1980)}]{Andreev1980}%
  \BibitemOpen
  \bibfield  {author} {\bibinfo {author} {\bibfnamefont {A.~F.}\ \bibnamefont
  {Andreev}},\ }\bibfield  {title} {\enquote {\bibinfo {title} {The
  hydrodynamics of two- and one-dimensional liquids},}\ }\href@noop {}
  {\bibfield  {journal} {\bibinfo  {journal} {Zh. Eksp. Teor. Fiz.}\ }\textbf
  {\bibinfo {volume} {78}},\ \bibinfo {pages} {2064} (\bibinfo {year}
  {1980})},\ \bibinfo {note} {[Sov. Phys. JETP {\bf 51}, 1038
  (1980)]}\BibitemShut {NoStop}%
\bibitem [{\citenamefont {Spohn}(2014)}]{Spohn2014}%
  \BibitemOpen
  \bibfield  {author} {\bibinfo {author} {\bibfnamefont {H.}~\bibnamefont
  {Spohn}},\ }\bibfield  {title} {\enquote {\bibinfo {title} {Nonlinear
  fluctuating hydrodynamics for anharmonic chains},}\ }\href@noop {} {\bibfield
   {journal} {\bibinfo  {journal} {J. Stat. Phys.}\ }\textbf {\bibinfo {volume}
  {154}},\ \bibinfo {pages} {1191} (\bibinfo {year} {2014})}\BibitemShut
  {NoStop}%
\bibitem [{\citenamefont {Kulkarni}\ and\ \citenamefont
  {Lamacraft}(2013)}]{KulkarniLamacraft2013}%
  \BibitemOpen
  \bibfield  {author} {\bibinfo {author} {\bibfnamefont {M.}~\bibnamefont
  {Kulkarni}}\ and\ \bibinfo {author} {\bibfnamefont {A.}~\bibnamefont
  {Lamacraft}},\ }\bibfield  {title} {\enquote {\bibinfo {title}
  {Finite-temperature dynamical structure factor of the one-dimensional {B}ose
  gas: {F}rom the {G}ross--{P}itaevskii equation to the
  {K}ardar--{P}arisi--{Z}hang universality class of dynamical critical
  phenomena},}\ }\href@noop {} {\bibfield  {journal} {\bibinfo  {journal}
  {Phys. Rev. A}\ }\textbf {\bibinfo {volume} {88}},\ \bibinfo {pages}
  {021603(R)} (\bibinfo {year} {2013})}\BibitemShut {NoStop}%
\bibitem [{\citenamefont {Benenti}\ \emph {et~al.}(2020)\citenamefont
  {Benenti}, \citenamefont {Lepri},\ and\ \citenamefont
  {Livi}}]{BenentiLepriLivi2020}%
  \BibitemOpen
  \bibfield  {author} {\bibinfo {author} {\bibfnamefont {G.}~\bibnamefont
  {Benenti}}, \bibinfo {author} {\bibfnamefont {S.}~\bibnamefont {Lepri}}, \
  and\ \bibinfo {author} {\bibfnamefont {R.}~\bibnamefont {Livi}},\ }\bibfield
  {title} {\enquote {\bibinfo {title} {Anomalous heat transport in classical
  many-body systems: {O}verview and perspectives},}\ }\href@noop {} {\bibfield
  {journal} {\bibinfo  {journal} {Front. Phys.}\ }\textbf {\bibinfo {volume}
  {8}},\ \bibinfo {pages} {292} (\bibinfo {year} {2020})}\BibitemShut {NoStop}%
\bibitem [{\citenamefont {Yamamoto}\ \emph {et~al.}(2006)\citenamefont
  {Yamamoto}, \citenamefont {Stopa}, \citenamefont {Tokura}, \citenamefont
  {Hirayama},\ and\ \citenamefont {Tarucha}}]{Yamamoto2006}%
  \BibitemOpen
  \bibfield  {author} {\bibinfo {author} {\bibfnamefont {M.}~\bibnamefont
  {Yamamoto}}, \bibinfo {author} {\bibfnamefont {M.}~\bibnamefont {Stopa}},
  \bibinfo {author} {\bibfnamefont {Y.}~\bibnamefont {Tokura}}, \bibinfo
  {author} {\bibfnamefont {Y.}~\bibnamefont {Hirayama}}, \ and\ \bibinfo
  {author} {\bibfnamefont {S.}~\bibnamefont {Tarucha}},\ }\bibfield  {title}
  {\enquote {\bibinfo {title} {Negative {C}oulomb drag in a one-dimensional
  wire},}\ }\href@noop {} {\bibfield  {journal} {\bibinfo  {journal} {Science}\
  }\textbf {\bibinfo {volume} {313}},\ \bibinfo {pages} {204} (\bibinfo {year}
  {2006})}\BibitemShut {NoStop}%
\bibitem [{\citenamefont {Debray}\ \emph {et~al.}(2001)\citenamefont {Debray},
  \citenamefont {Zverev}, \citenamefont {Raichev}, \citenamefont {Klesse},
  \citenamefont {Vasilopoulos},\ and\ \citenamefont {Newrock}}]{Debray2001}%
  \BibitemOpen
  \bibfield  {author} {\bibinfo {author} {\bibfnamefont {P.}~\bibnamefont
  {Debray}}, \bibinfo {author} {\bibfnamefont {V.}~\bibnamefont {Zverev}},
  \bibinfo {author} {\bibfnamefont {O.}~\bibnamefont {Raichev}}, \bibinfo
  {author} {\bibfnamefont {R.}~\bibnamefont {Klesse}}, \bibinfo {author}
  {\bibfnamefont {P.}~\bibnamefont {Vasilopoulos}}, \ and\ \bibinfo {author}
  {\bibfnamefont {R.~S.}\ \bibnamefont {Newrock}},\ }\bibfield  {title}
  {\enquote {\bibinfo {title} {Experimental studies of {C}oulomb drag between
  ballistic quantum wires},}\ }\href@noop {} {\bibfield  {journal} {\bibinfo
  {journal} {J. Phys.: Condens. Matter}\ }\textbf {\bibinfo {volume} {13}},\
  \bibinfo {pages} {3389} (\bibinfo {year} {2001})}\BibitemShut {NoStop}%
\end{thebibliography}%

\end{document}